\documentclass[article]{emulateapj}

\newcommand{\co}{\mbox{\rm CO}}

\newcommand{\hi}{\mbox{\rm \ion{H}{1}}}

\newcommand{\xcounits}{\mbox{cm$^{-2}$ (K km s$^{-1}$)$^{-1}$}}

\newcommand{\xco}{\mbox{$X_{\rm CO}$}}

\newcommand{\dustmass}{\mbox{$3 \times 10^5$~M$_{\odot}$}}
\newcommand{\htwomass}{\mbox{$3.2 \times 10^7$~M$_{\odot}$}}
\newcommand{\colddustmass}{\mbox{$2 \times 10^6$~M$_{\odot}$}}

\newcommand{\logdtog}{\mbox{$-2.78$}}
\newcommand{\dtograt}{\mbox{$600$}}
\newcommand{\logdtogwhtwo}{\mbox{$-2.86$}}
\newcommand{\logdtogwhe}{\mbox{$-3.0$}}
\newcommand{\dtogwhtworat}{\mbox{$700$}}
\newcommand{\logdtogwing}{\mbox{$-2.94$}}
\newcommand{\logdtogbar}{\mbox{$-2.71$}}
\newcommand{\logdtogco}{\mbox{$-2.54$}}

\newcommand{\extfact}{\mbox{1.3}}
\newcommand{\intxco}{\mbox{$13 \pm 1 \times 10^{21}$}}
\newcommand{\peakxco}{\mbox{$6 \pm 1 \times 10^{21}$}}

\shorttitle{S3MC: FIR in the SMC} 
\shortauthors{Leroy et al.}
\slugcomment{Accepted for publication in \emph{The Astrophysical Journal}}

\begin{document}
\title{The Spitzer Survey of the Small Magellanic Cloud: FIR Emission
and Cold Gas in the SMC}

\author{Adam Leroy\altaffilmark{1,2}, Alberto Bolatto\altaffilmark{2}, Snezana
  Stanimirovic\altaffilmark{3,2}, Norikazu Mizuno\altaffilmark{4}, Frank
  Israel\altaffilmark{5}, and Caroline Bot\altaffilmark{6}}

\altaffiltext{1}{Max-Planck-Institut f\"{u}r Astronomie, Königstuhl 17, D-69117,
  Heidelberg, Germany}

\altaffiltext{2}{Radio Astronomy Lab, UC Berkeley, 601 Campbell Hall,
  Berkeley, CA, 94720}

\altaffiltext{3}{Department of Astronomy, University of Wisconsin, 475 North
  Charter Street, Madison, WI, 53706}

\altaffiltext{4}{Department of Astrophysics, Nagoya University, Furo-cho,
  Chikusa-ku, Nagoya 464-8602, Japan}

\altaffiltext{5}{Sterrewacht Leiden, PO Box 9513, 2300 RA Leiden, The
  Netherlands}

\altaffiltext{6}{Spitzer Science Center, California Institute of Technology,
  1200 East California Boulevard, Pasadena, CA 91125}

\begin{abstract}
  We present new far infrared maps of the Small Magellanic Cloud (SMC) at 24,
  70, and 160~$\mu$m obtained as part of the {\em Spitzer} Survey of the Small
  Magellanic Cloud \citep[S$^3$MC,][]{BOLATTO06}. These maps cover most of the
  active star formation in the SMC Bar and the more quiescent Wing. We combine
  our maps with literature data to derive the surface density across the SMC.
  We find a total dust mass of $M_{dust} = $\dustmass, implying a
  dust-to-hydrogen ratio over the region studied of $\log_{10} D/H =
  \logdtogwhtwo$, or 1-to-\dtogwhtworat, which includes H$_2$. Assuming the
  dust to trace the total gas column, we derive H$_2$ surface densities across
  the SMC. We find a total H$_2$ mass $M_{\rm{H2}} =$ \htwomass\ in a
  distribution similar to that of the CO, but more extended.  We compare
  profiles of CO and H$_2$ around six molecular peaks and find that on average
  H$_2$ is more extended than CO by a factor of $\sim \extfact$.  The implied
  CO-to-H$_2$ conversion factor over the whole SMC is $\xco = \intxco$
  \xcounits. Over the volume occupied by CO we find a lower conversion factor,
  $\xco = \peakxco$ \xcounits, which is still a few times larger than that
  found using virial mass methods. The molecular peaks have H$_2$ surface
  densities $\Sigma_{\rm{H2}} \approx 180 \pm 30$ M$_{\odot}$ pc$^{-2}$,
  similar to those in Milky Way GMCs, and correspondingly low extinctions,
  $A_V \sim 1$ -- $2$ mag. To reconcile these measurements with predictions by
  the theory of photoionization-regulated star formation, which requires $A_V
  \sim 6$, the GMCs must be $\sim 3$ times smaller than our 46 pc resolution
  element.  We find that for a given hydrostatic gas pressure, the SMC has a
  $2$ -- $3$ times lower ratio of molecular to atomic gas than spiral
  galaxies. Combined with the lower mean densities in the SMC this may explain
  why this galaxy has only 10\% of its gas in the molecular phase.
\end{abstract}

\keywords{ISM: dust, ISM: molecules, Galaxies:Magellanic Clouds,
Galaxies:ISM, Infrared: ISM}

\section{Introduction}

Dust grains play a central role in the formation of stars from atomic gas.
Dust shields molecular gas from dissociating radiation and allows it to cool
and condense to the densities necessary to form stars. The surfaces of dust
grains serve as the sites of molecular hydrogen formation. In low metallicity
galaxies, an underabundance of dust may result in more intense radiation
fields, different thermal balance in the interstellar medium (ISM), and
different structure for giant molecular clouds (GMCs). However, there are only
a few low metallicity systems in which atomic gas, molecular gas, dust, and
star formation tracers have all been mapped at resolution sufficient to
resolve individual star-forming regions or GMCs ($\lesssim 100$ pc).

In this paper we present new far infrared (FIR) images of the Small Magellanic
Cloud (SMC) obtained using the {\em Spitzer} Space Telescope as part of the
{\em Spitzer} Survey of the Small Magellanic Cloud
\citep[S$^3$MC,][]{BOLATTO06}. We combine these maps with previous FIR, \hi\,
and CO data to construct a complete picture of the star-forming ISM in this
nearby low metallicity galaxy. We focus on three questions: 1) what is the
mass of dust in the SMC?, 2) how much molecular hydrogen (H$_2$) is in the
SMC? and 3) how does the distribution of H$_2$ in the SMC --- particularly its
structure and relation to CO emission --- compare to that in the Milky Way?
In this introduction we describe the SMC and expand on these questions.

The SMC is the nearest actively star-forming, low metallicity ($Z~\lesssim
1/10~Z_\odot$) galaxy. It is therefore good location to study the relationship
between gas, dust, and star formation at low metallicity. It forms stars at
$\sim 0.05$ M$_{\odot}$ yr$^{-1}$ \citep{WILKE04} --- a normalized rate that
is comparable to that of the Milky Way \citep[the dynamical mass of the SMC is
$\sim 2 \times 10^9$ M$_{\odot}$---][]{STANIMIROVIC04}. It is gas rich, with
$\approx 4 \times 10^8$ M$_{\odot}$ of atomic gas in an extended distribution
\citep{STANIMIROVIC99} and a comparable mass of stars, $\sim 3 \times 10^8$
M$_{\odot}$ (estimated from the DIRBE 2.2~$\mu$m data). The SMC is also
unenriched in heavy elements, with an oxygen abundance just above a tenth
solar \citep[$12 + \log \left( \rm{O/H} \right) = 8.0$,][]{DUFOUR84}. At a
distance of 61.1 kpc \citep{HILDITCH05,KELLER06}, even modest telescopes can
observe the SMC with excellent spatial resolution, and at a Galactic latitude
of $b = -44^\circ$ it is obscured by little foreground material.

Several recent studies have used FIR emission to study the dust in the SMC,
finding evidence for a low dust-to-gas ratio, $DGR$.  \citet{STANIMIROVIC00}
modeled HIRAS data (a high resolution IRAS product) and found that the SMC had
an unexpectedly low dust mass of only $1.8 \times 10^4$ M$_{\odot}$. The
corresponding $DGR$ is two orders of magnitude below the Galactic value.
\citet{BOT04} studied the diffuse gas using ISO and IRAS data, and found that
the emissivity of diffuse SMC gas is $30$ times lower than that of high
latitude Galactic gas. \citet{WILKE04} studied emission integrated over the
whole SMC, also using ISO data. They fit the dust using several black bodies
and found a much higher total dust content than previous studies, $7.8 \times
10^5$ M$_{\odot}$. Using this result, they derived a $DGR$ only a few times
lower than the Galactic value.

Only $1.7 \times 10^5$ M$_{\odot}$ of the dust mass obtained by
\citet{WILKE04} would be estimated from their FIR data alone. The rest is in a
cold component that emits at millimeter wavelengths, with a temperature of
$\sim 10$~K. \citet{STANIMIROVIC00} also recognized the possibility of a large
mass of cold dust from the 140 and 240 $\mu$m DIRBE data, which show higher
intensities than those allowed by a single population of Galactic dust that
fits the IRAS measurements.  \citet{AGUIRRE03} demonstrated that this excess
extends out to 1.2 mm by measuring the spectrum of the SMC using the TopHat
balloon. They also showed, however, that the measurements can be fit by a
single dust population if $\beta = 1$ for the dust emissivity law,
$\tau\propto\lambda^{-\beta}$.

In addition to providing information about the dust, the FIR emission can be
used to trace molecular gas. Clouds of molecular hydrogen are the sites of
star formation, but H$_2$ cannot be observed directly under the conditions
found in GMCs because it lacks a permanent dipole moment and $T$ is too low to
excite the quadrupole transitions. Observations of tracer molecules,
particularly CO, often substitute for direct observations of H$_2$.  In low
metallicity galaxies, the lack of heavy elements and the lower $DGR$ may alter
the structure of molecular clouds and affect the relationship between tracer
molecules and H$_2$, leaving the amount of molecular gas in these systems
uncertain. Unlike the CO-to-H$_2$ ratio, the dust-to-hydrogen ratio, $D/H$,
can be measured away from molecular clouds using \hi\ and FIR emission. Under
the assumption that $D/H$ varies slowly, FIR emission may provide a better
estimate of the amount of H$_2$ in these systems than CO observations.

The molecular hydrogen content of the SMC may be even less well known
than its dust content.  \citet{MIZUNO01,MIZUNO06} surveyed the CO
J$=1\rightarrow0$ line across most of the SMC using the NANTEN
telescope and found a luminosity of $\sim 1 \times 10^5$ K km s$^{-1}$
pc$^2$. They used the virial mass method to derive a CO-to-H$_2$
conversion factor, $\xco$, of 1--5 $\times 10^{21}$ \xcounits, about
10 times the Galactic value ($\xco^{Gal}\approx2 \times
10^{20}$~\xcounits). This implies a total molecular mass of $M_{\rm{mol}}
\sim 2$ -- $10 \times 10^6$ M$_{\sun}$. \citet{BLITZ06} corrected
these data for resolution and sensitivity effects and found a value of
$\xco$ towards the low end of this range, $1$ -- $1.5 \times
10^{21}$ \xcounits, implying $M_{\rm{mol}} \sim 2$--$3 \times 10^6$
M$_{\sun}$. Considering individual regions with better spatial
resolution, \citet{RUBIO93} and \citet{BOLATTO03} both found
conversion factors a few times the Galactic value, though dependent on
the size of the structure studied.

Modeling of the FIR can be used to derive the mass and structure of H$_2$
without relying on the assumptions that CO and H$_2$ have the same structure
and that GMCs are in virial equilibrium. If the gas and dust are well mixed
and properties do not vary too dramatically between the molecular and atomic
ISM, the dust acts as an optically thin tracer of the total, \hi$+$H$_2$, gas
surface density.  \citet{DAME01} demonstrated this method to work comparing
IRAS and CO in the Milky Way \citep[see also][]{REACH98}. \citet{ISRAEL97}
used IRAS to study the SMC and found a total H$_2$ content of $7.5 \times
10^7$ M$_{\odot}$ and the CO-to-H$_2$ conversion factor to be 60 times the
Galactic value. In this paper we use a similar method to construct H$_2$ maps
with $46$ pc resolution and sensitivity to H$_2$ surface densities as low as
$\sim 20$~M$_{\odot}$~pc$^{-2}$ (3$\sigma$).

Using the H$_2$ map, we test three predictions regarding molecular gas in low
metallicity systems. \citet{MCKEE89} argued that photoionization by far
ultraviolet photons sets the level of ionization inside GMCs and thus
regulates their structure. Magnetic fields couple to the ionized gas and
support cores against collapse. Thus, the level of ionization is tied to the
rate of star formation in the cloud. Low mass star formation provides feedback
that keeps the mean extinction, and with it the ionization level, across the
cloud at a particular value. In the Milky Way this value is $\bar{A_V} \sim 4$
-- $8$, independent of cloud mass.  In low metallicity systems, with low
$DGR$s, \citet{MCKEE89} predicts GMCs will require higher gas surface
densities, $\Sigma_{\rm{HI + H2}}$, in order to achieve the same mean
extinctions found in the Milky Way. For the low $DGR$ in the SMC, this implies
GMCs with $\Sigma_{\rm{HI + H2}} \gtrsim 1000$ M$_{\odot}$ pc$^{-2}$
\citep[see also][]{PAK98}.  Second, CO in the SMC may be underabundant or
absent in regions where H$_2$ survives because the H$_2$ self-shields but this
does not stop photodissociation of the CO \citep[e.g.,][]{MALONEY88}. Third,
\citet{BLITZ04} showed that there is a good correlation between the midplane
hydrostatic pressure and the ratio of molecular to atomic gas along lines of
sight towards spiral galaxies.  We test whether this correlation extends to
the low metallicity, low pressure environment found in the SMC, or whether
metallicity acts as a ``third parameter'' in the relationship between pressure
and molecular gas fraction.

This paper is organized as follows: in \S \ref{DATA} we discuss the new {\em
  Spitzer} maps and the literature data we use to supplement them. We describe
our foreground subtraction and discuss uncertainties in the data.  In \S
\ref{RESULTS} we present the new maps and literature data on a common
astrometric grid. We construct maps of dust and H$_2$ and present an updated
FIR spectral energy distribution (SED). In \S \ref{ANALYSIS} we consider the
relationship between dust and gas, including the dust-to-gas ratio, the
CO-to-H$_2$ conversion factor, the structure of molecular clouds, and pressure
and extinction in the star-forming regions. In \S \ref{CONCLUSIONS} we present
our conclusions.

\section{Data}
\label{DATA}

To meet the science goals described in the introduction, we require
FIR images of the SMC at several wavelengths on a common astrometric
grid with good relative calibration. These allow us to derive maps of
the dust and H$_2$ content in \S~\ref{RESULTS} and add new points at
24, 70, and 160 $\mu$m to the integrated SED of the SMC. We start with
the new {\em Spitzer} images of the SMC from S$^3$MC at 24, 70, and
160 $\mu$m and supplement these data with images at 60 and 100 $\mu$m
from IRAS and at 170 $\mu$m from ISO. We use data from DIRBE from 25
-- 240 $\mu$m as the basis for the flux calibration and consider
TopHat data \citep{AGUIRRE03} when constructing the integrated
SED. The wavelengths, telescopes, limiting resolution, and references
for the data used in this paper are listed in Table~\ref{DATATAB}. In
the rest of this section we describe how we isolate the emission from
the SMC in the FIR images and estimate the uncertainties in the final
maps.

\begin{deluxetable*}{l l c c l}
\tabletypesize{\small}
\tablewidth{0pt}
\tablecolumns{5}
\tablecaption{\label{DATATAB} Data Sets Used in This Paper}

\tablehead{\colhead{Nominal Wavelength} & \colhead{Data Set} &
\colhead{Limiting} & \colhead{Spatial} & \colhead{Reference} \\~~($\mu$m) & & Resolution & Resolution\tablenotemark{a} & }

\startdata
24, 70, 160 & MIPS ({\it Spitzer}) & $40\arcsec$ & 12~pc & \citet[][]{BOLATTO06} \\
25, 60, 100, 140, 240\tablenotemark{b} & DIRBE (COBE) & $0^{\circ}.7$ & 750~pc & \citet[][]{HAUSER98} \\
25, 60, 100 & HIRAS (IRAS) & $1\arcmin.7$ & 30~pc & \citet[][]{STANIMIROVIC00} \\
25, 60, 100 & IRIS (IRAS) & $4\arcmin.3$ & 76~pc & \citet[][]{IRISPAPER} \\
170 & ISOPHOT (ISO) & $1\arcmin.5$ & 27~pc & \citet[][]{BOT04} \\
\hi\ & ATCA + Parkes & $1\arcmin.5$ & 27~pc & \citet{STANIMIROVIC99} \\
CO & NANTEN & $2\arcmin.6$ & 46~pc& \citet{MIZUNO01,MIZUNO06}
\enddata
\tablenotetext{a}{At our adopted distance of 61.1~kpc \citep{HILDITCH05,KELLER06}.}
\tablenotetext{b}{The DIRBE zodiacal light subtracted mission average (ZSMA) data are the basis of our photometric
calibration in this paper.}
\end{deluxetable*}

\subsection{Color Corrections and Flux Definitions}

Throughout this paper, we quote monochromatic flux densities assuming
a $F_{\nu} \propto \nu^{-1}$ spectral energy distribution across the
relevant bandpass. This is the convention for IRAS, IRAC, and DIRBE,
but differs from the MIPS standard, which is to assume a
Rayleigh-Jeans tail ($F_{\nu} \propto \nu^2$) across the bandpass. We
apply color corrections to the MIPS data at $24$, $70$, and $160$
$\mu$m by dividing by $0.961$, $0.918$, and $0.959$, respectively,
before any processing to bring them to the IRAS definition. For most
of the work in this paper, this choice is arbitrary.

\subsubsection{Bright Pixel Removal}

Before other processing we remove pixels with unexpectedly high
intensities from the 24, 70, and 160 $\mu$m maps ($>100$, $>200$, and
$>250$ MJy ster$^{-1}$, respectively). At 24 $\mu$m, these pixels
correspond to point sources and are well over the saturation limit, at
160 $\mu$m these are simply artifacts, and at 70 $\mu$m they are a mix
of the two. Removing these point sources lowers the integrated flux in
the foreground subtracted 24 $\mu$m map by $\approx 15\%$. The effect
on the flux at 70 $\mu$m is negligible ($\sim 1$\%). We interpolate to
fill these pixels using nearby data when placing the maps at a common
resolution by taking the beam-weighted mean of all nearby unflagged
data.

\subsection{Foreground Removal}

The intensity in the FIR maps is the sum of emission from the source
and several foregrounds,

\begin{equation}
I_{\mbox{FIR}} = I_{\mbox{Zod}} + I_{\mbox{Gal+CIB}} +
I_{\mbox{Source}}~.
\end{equation}

\noindent $I_{Zod}$ is the intensity of zodiacal light, which
dominates the foreground at shorter wavelengths ($12$ -- $70~\mu$m).
$I_{Gal+CIB}$ is Galactic dust emission and a small diffuse cosmic
infrared background term. This is the primary foreground at longer
wavelengths ($100$ -- $240$ $\mu$m).

\subsubsection{Zodiacal Light}
\label{ZODSECT}

The HIRAS, IRIS, ISO, and DIRBE data have all been corrected for
zodiacal light. We estimate the contribution of zodiacal light to the
MIPS maps using the model of interplanetary dust by \citet{KELSALL98}
implemented in the {\em Spitzer} Observation Planning Tool. A planar
fit derived to this model near the SMC for our observations is:

\begin{equation}
I_{Zod} = I_{0,Zod}~\left[0.006 (l - l_{SMC}) - 0.003 (b - b_{SMC}) + 1\right]
\end{equation}

\noindent where $I_{0,Zod}$ is the intensity of zodiacal light at the
nominal center of the SMC ($l = 302^\circ.7969$, $b = -44^\circ.2992$)
on the date of the observations. At 24, 70, and 160 $\mu$m,
respectively, $I_{0,Zod}$ is $20.1$, $5.3$, and $0.9$ MJy ster$^{-1}$
($98\%$, $36\%$, and $6\%$ of the total flux, respectively). The
variation in $I_{Zod}$ over the several days during which the
observations were taken is minimal.

\subsubsection{Galactic Foreground and Cosmic Infrared Background}

At 100 $\mu$m and longer wavelengths the Galactic ISM contributes the
largest foreground. We apply a similar treatment to \citet{BOT04}, and
estimate the Galactic foreground along lines of sight towards the SMC
using the Galactic \hi\ column density, which is clearly separated
from the SMC \hi\ column density by its velocity. We use the Galactic
\hi\ column density observed at $\sim 0.5^{\circ}$ resolution from the
southern survey by \citet[][]{BAJAJA05}. For the IRIS $100$ $\mu$m
map, we use the fit of FIR intensity to \hi\ column density by
\citet{BOULANGER96}. At 160~$\mu$m and 170~$\mu$m, we interpolate from
the fits at $140$ and $240$~$\mu$m by \citet{BOULANGER96} into these
bands using their best fit temperature $T = 17.5$~K and emissivity
$\beta = 2$. The median intensity of the Galactic foreground plus
cosmic infrared background (CIB) is $1.8$, $4.7$, and $4.8$ MJy
ster$^{-1}$ at 100, 160, and 170 $\mu$m, respectively.

For the 60 and 70 $\mu$m maps, the foreground due to Galactic dust is
small, but \citet{BOULANGER96} do not provide fits for these bands. We
use estimates of the emission of local dust per unit \hi\ column
density at 60~$\mu$m from \citet{DESERT90} and interpolate to
70~$\mu$m using the average colors in the map. This calculation is
approximate, but the median intensity of this foreground at these
wavelengths is small: 0.5 MJy ster$^{-1}$ and 1 MJy ster$^{-1}$, or
$\lesssim 20\%$ of the zodiacal light foreground.

We do not subtract any foregrounds from the HIRAS maps; \citet{STANIMIROVIC00}
define regions of sky near the SMC as blank and subtract a single foreground
that includes zodiacal light, the Galactic foreground, and the CIB. Because we
do not have an independent estimate of the zodiacal light for these data, we
cannot break this foreground into its individual components. Instead, we bring
the HIRAS data onto a consistent foreground subtraction by comparing them to
the foreground subtracted IRIS data at a common resolution. We find that
offsets of 0.75 MJy ster$^{-1}$ and 1.5 MJy ster$^{-1}$ at 60 and 100 $\mu$m,
respectively, are needed to bring the HIRAS zero level into agreement with
IRIS. After this correction, the agreement between the two IRAS data sets is
excellent.

\subsubsection{Foreground Subtracting the DIRBE Data}

The DIRBE Zodiacal-Subtracted Mission Average (ZSMA) maps appear to
have a small amount of residual zodiacal light, $\sim1$~MJy
ster$^{-1}$ judging from a plot of intensity as a function of \hi\
column over the whole sky. Therefore, we treat the DIRBE foreground
subtraction separately. We derive DIRBE intensities as a function of
\hi\ column density near the SMC (within $10^{\circ}$ of the optical
center) but offset enough that we see only Galactic FIR emission (more
than $3^{\circ}$ away from the optical center). Then, for lines of
sight towards the SMC we measure the Milky Way \hi\ column density as
above and subtract the average intensity of lines of sight with the
same \hi\ column density away from the SMC. This is our estimate of
the DIRBE foreground, which should account for Galactic emission, the
CIB, and residual zodiacal light.

The DIRBE 25 $\mu$m data is overwhelmingly contaminated by zodiacal
light. We fit a plane to the nearby data and subtract this foreground
before using it.

\subsubsection{The MIPS 24 $\mu$m Map Foreground}

In the case of the 24 $\mu$m map there are large regions of the map free of
diffuse emission. Because the zodiacal light foreground is large and
potentially uncertain at 24~$\mu$m and we do not have an estimate of the
Galactic foreground, we use blank regions in the {\em Spitzer} map to estimate
the residual foreground. We measure an additional foreground with a median
value of 2.5 MJy ster$^{-1}$ above and beyond the original zodiacal light
foreground (\S \ref{ZODSECT}). This value is probably too large to be due to
the to Galactic foreground alone and may represent a failure in the zodiacal
light model or a calibration error for MIPS. The presence of blank sky in the
24 $\mu$m map makes it easy to remove, regardless.

\subsubsection{Consistency and Calibration}

We check the calibration of the high resolution data against DIRBE and find
generally good agreement. The fluxes from each map integrated over the whole
SMC agree (see Table \ref{FLUXTAB}). The exception is the MIPS 160 $\mu$m map,
which contains substantially more flux than we estimate based on the DIRBE
data. A central goal of the DIRBE mission was to obtained absolute brightness
maps of the sky \citep{HAUSER98}, and that data represents a more reliable
measurement of the flux than our MIPS map. Therefore, we calibrate the 160
$\mu$m to match DIRBE by rescaling the map. Appendix \ref{CALMIPSAPP}
describes how we carry out this calibration in detail. The MIPS 70 $\mu$m map
is consistent with the data at 60 $\mu$m and 100 $\mu$m, but because the SED
has a very steep slope over the MIPS 70 $\mu$m bandpass, this consistency is
not a particularly strong constraint. It is not possible to perform the same
sort of detailed calibration we adopt at 160 $\mu$m for the 70 $\mu$m map.

\subsubsection{Saturation}
\label{SATSECT}

Saturation may represent a concern over part of our 160 $\mu$m map. The bright
star-forming region in the southwest of the galaxy contains a significant
region with brightness greater than 50 MJy ster$^{-1}$, the nominal saturation
limit for our medium speed scan maps. A few other star-forming regions (N~76
to the north and N~83 in the east) also contain small regions with intensities
above 50 MJy ster$^{-1}$.  In total, 13\% of the flux and 2.5\% of the area
are in regions with intensities above the saturation limit. In these regions,
the response of detector becomes nonlinear and we have an additional gain
uncertainty. The figure in Appendix A shows that in the southwest of the bar
--- the source of the lines of sight with intensities $\geq 25$ MJy
ster$^{-1}$ at the DIRBE resolution --- the MIPS data are $\sim 20$ -- $30\%$
lower than the DIRBE data. Outside this region, the MIPS and DIRBE data are
well matched after correction.

\subsection{Uncertainties in the Final Maps}

We estimate the statistical uncertainties for the FIR data from the scatter
about the median in regions of low intensity emission.  We fit a Gaussian to
the residuals about the local median taken over at $4' \times 4'$ region, so
that for each point $I_{resid} = I_{map} - I_{median}~(\pm2\arcmin)$ with
$I_{median}~(\pm2\arcmin)$ the median over a $4\arcmin \times 4\arcmin$ box
centered on that point. We conduct these measurements in regions of low
intensity emission outside the main body of the galaxy to minimize the effect
of structure on the measurement. However, emission with small scale structure
and artifacts present in the MIPS and ISO maps contribute to the noise values.
These are legitimate sources of uncertainty, because they affect measurements.

\begin{figure}
\begin{center}
\epsscale{1.0}
\plotone{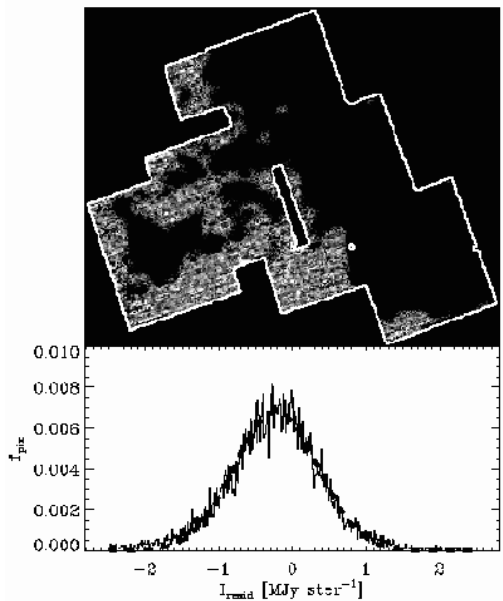}
 
\figcaption{\label{MIPSNOISE} (top) The map used to measure the noise at 160
  $\mu$m. The map shows fluctuations of the 160 $\mu$m MIPS map about the
  local median (over a $4' \times 4'$ box centered on the pixel). Regions with
  bright emission or no data are blanked out and appear black. Artifacts,
  particularly striping along the direction of the map, contribute to the
  noise, but very little extended structure is visible. The extent of the MIPS
  160 $\mu$m map is shown by the white border. (below) The histogram of the
  residuals about the median. The dashed lines shows the best fit Gaussian,
  which has $\sigma_{160} = 0.58$ MJy ster$^{-1}$.}

\end{center}
\end{figure}

Table \ref{NOISETAB} gives the results of our noise estimates, and Figure
\ref{MIPSNOISE} shows a map and histogram of deviations about the median for
the MIPS 160 $\mu$m data. Two artifacts of small scale structure are visible:
the tail of low, but non-zero, values towards negative intensities and the
slight shift of the center of the Gaussian towards negative intensities. Table
\ref{NOISETAB} gives the noise for each map at the native and IRIS resolution.
For most maps, the noise scales approximately as expected with averaging.
Deviations from this scaling may be the result of artifacts such as the
visible striping at low intensity in the MIPS data.

Because we calibrate our data against DIRBE, the absolute
uncertainties in DIRBE are relevant to our data. \citet{HAUSER98}
gives the absolute gain uncertainty for DIRBE at $25$, $60$, $100$,
$140$, and $240$ $\mu$m as 15.1, 10.4, 13.5, 10.6, and 11.6 \%,
respectively. The uncertainties in the absolute offset of the DIRBE
data are all small enough to be negligible for our purposes. As a
result of gain uncertainties for DIRBE, the minimum systematic
uncertainty in any absolute flux or intensity in this paper is $\sim
10\%$.

The systematic uncertainty associated with the foreground subtraction is $\pm
1$ MJy ster$^{-1}$. This is the approximate magnitude of the scatter about our
best fits of DIRBE intensity to \hi\ column density ($\sim 0.2$ MJy
ster$^{-1}$ at 100 $\mu$m, $\sim 1.2$ MJy ster$^{-1}$ at 240 $\mu$m); it is
also about the magnitude of the offset needed to bring IRIS and HIRAS into
rough agreement ($\sim 0.75$ and $1.5$ MJy ster$^{-1}$).

\begin{deluxetable}{l c l}
\tabletypesize{\small}
\tablewidth{0pt}
\tablecolumns{5}
\tablecaption{\label{NOISETAB} Random Uncertainties in FIR Maps}

\tablehead{\colhead{Data Set} & \colhead{$\sigma$} & \colhead{$\sigma$} \\ 
&  Native Resolution\tablenotemark{a} & IRIS
  Resolution\tablenotemark{b} \\
& MJy ster$^{-1}$ & MJy ster$^{-1}$ }

\startdata
24 $\mu$m\tablenotemark{a} MIPS ({\it Spitzer}) & $ 0.02$ & $ 0.01$ \\
60 $\mu$m HIRAS (IRAS) & $ 0.07$ & $ 0.02$ \\
60 $\mu$m IRIS (IRAS) & $ 0.02$ & $ 0.02$ \\
70 $\mu$m\tablenotemark{a} MIPS ({\it Spitzer}) & $ 0.34$ & $ 0.06$ \\
100 $\mu$m HIRAS (IRAS) & $ 0.10$ & $ 0.05$ \\
100 $\mu$m IRIS (IRAS) & $ 0.05$ & $ 0.05$ \\
160 $\mu$m MIPS ({\it Spitzer}) & $0.58$ & $0.10$ \\
170 $\mu$m ISOPHOT (ISO) & $ 0.22$ & $0.05$
\enddata

\tablenotetext{a}{24 and 70 $\mu$m data convolved to 160 $\mu$m resolution.}

\tablenotetext{b}{Uncertainty after convolution to the IRIS ($4'$)
resolution.}
\end{deluxetable}

\section{Results}
\label{RESULTS}

In this section we present maps of FIR emission and neutral gas from
the SMC. We use these maps to derive the dust and molecular gas
surface density and we present an updated SED of the SMC that includes
the new 24, 70, and 160 $\mu$m data.

\subsection{FIR Maps of the SMC}

\begin{figure*}
\begin{center}
\plotone{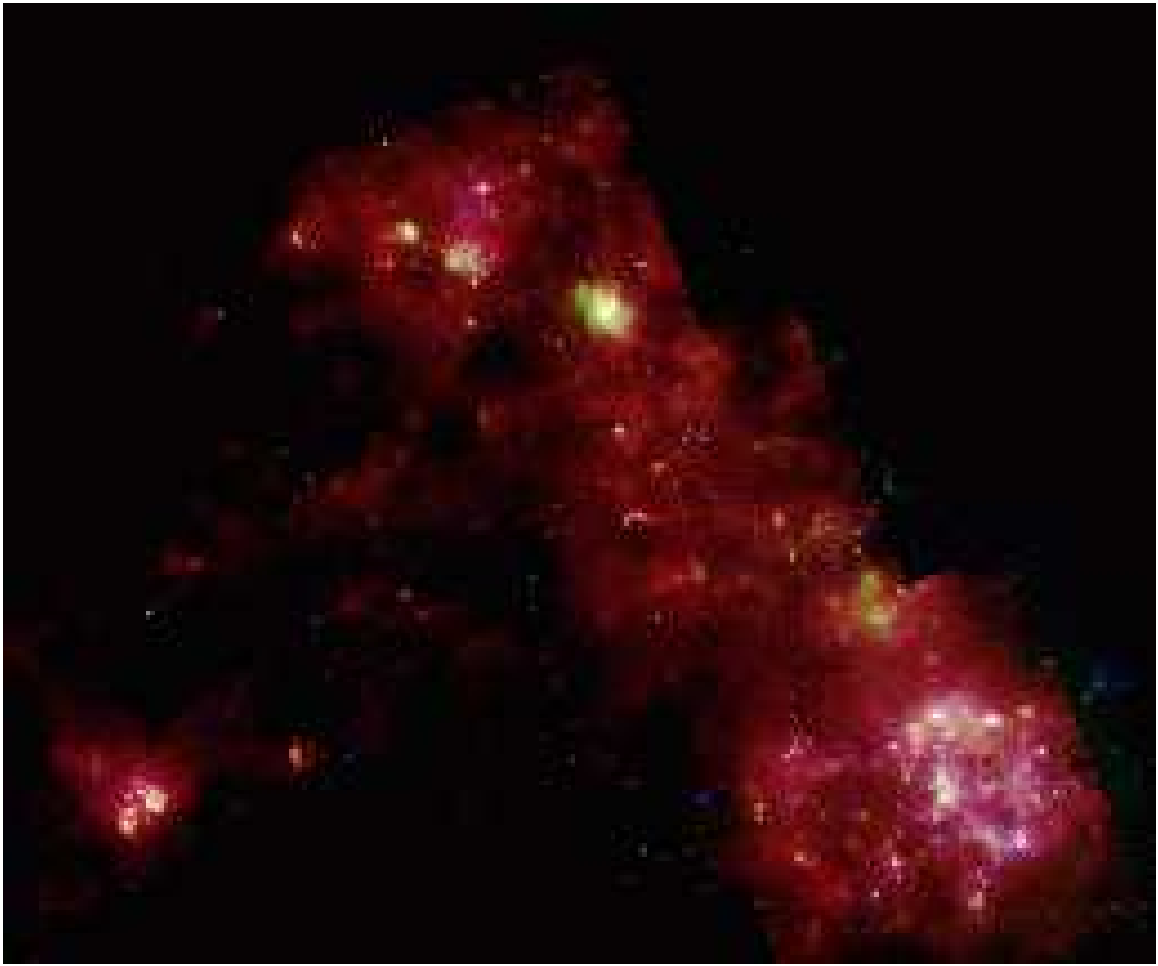} 

\figcaption{\label{RGB} A three color image showing the S$^3$MC maps
  \citep{BOLATTO06} of the SMC at 160 $\mu$m (red), 24 $\mu$m (green), and 8
  $\mu$m (blue). The 160 $\mu$m emission comes from big grains and extends
  across the whole SMC. Diffuse 24 $\mu$m emission, from hot very small grains
  (VSGs), and the 8 $\mu$m emission, from PAHs, are both confined to the
  regions of active star formation.}

\end{center}
\end{figure*}

Figure \ref{RGB} shows the relative extent of 160 $\mu$m (red), 24
$\mu$m (green), and 8 $\mu$m (blue) emission in the S$^3$MC maps
\citep{BOLATTO06}. The 160 $\mu$m emission is bright over the whole
SMC Bar in the west and in the N~83/N~84 star-forming region far to
the east. This emission comes mostly from big grains ($a \approx 15$
-- $110$~nm) with $T \sim 20$~K with a small (few percent)
contribution from the [CII] 158 $\mu$m line (see Appendix
\ref{CALMIPSAPP}). Every region in the S$^3$MC map shows some 160
$\mu$m emission including the SMC Wing, which is roughly
perpendicular to and east of the bar and contains more diffuse gas. By
contrast, the 24 $\mu$m emission and the 8 $\mu$m emission are both
confined to regions of active star formation and point sources. Much
of the 24 $\mu$m emission comes from stochastically heated very small
grains, though very hot ($\sim 100$~K) large grains may also make a
contribution. In environments similar to the Solar Neighborhood, most
strong 8$\mu$m emission appears to originate from polycyclic aromatic
hydrocarbons. However, very small grains may produce emission in this
band even if PAHs are absent. All three bands are brightest towards
the N~66 star-forming region in the southwest part of the galaxy. This
is the brightest star-forming region in the SMC, home to $\sim 60$ O
stars \citep{MASSEY89}.

\begin{figure*}
\begin{center}
\epsscale{1.0}
\plottwo{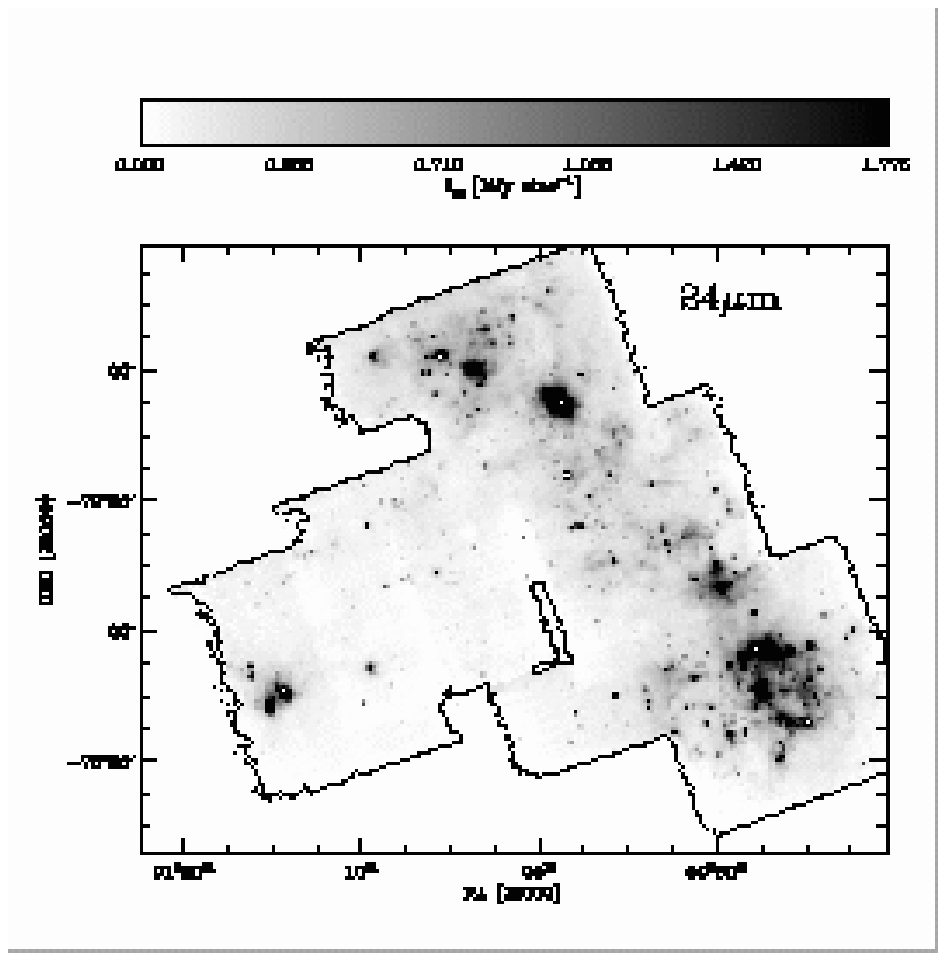}{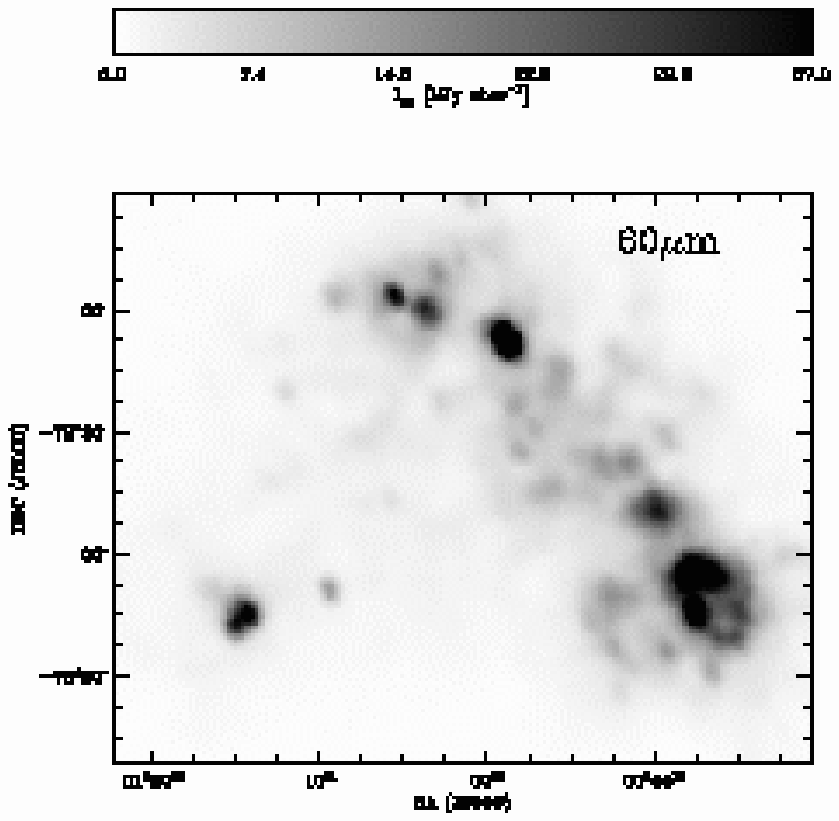}
\plottwo{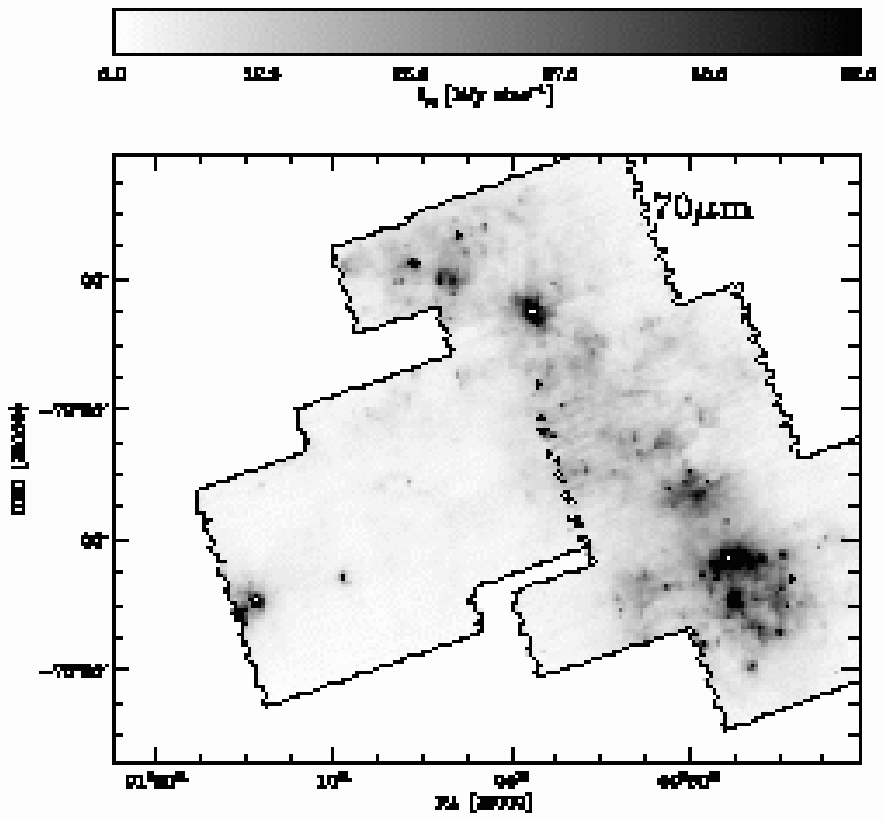}{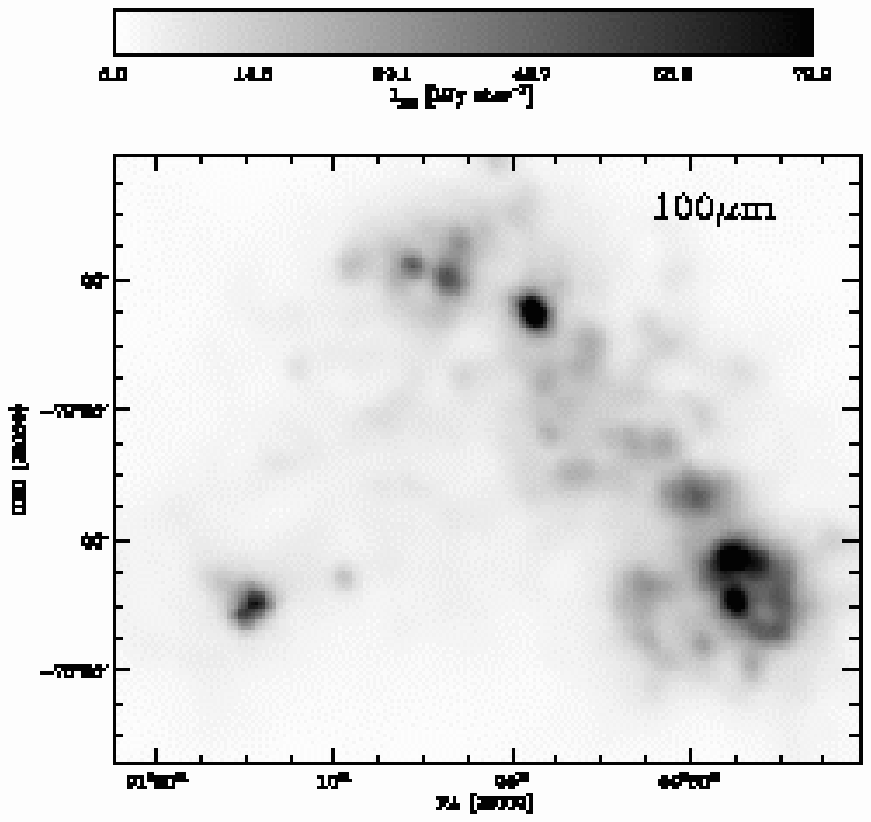}
\plottwo{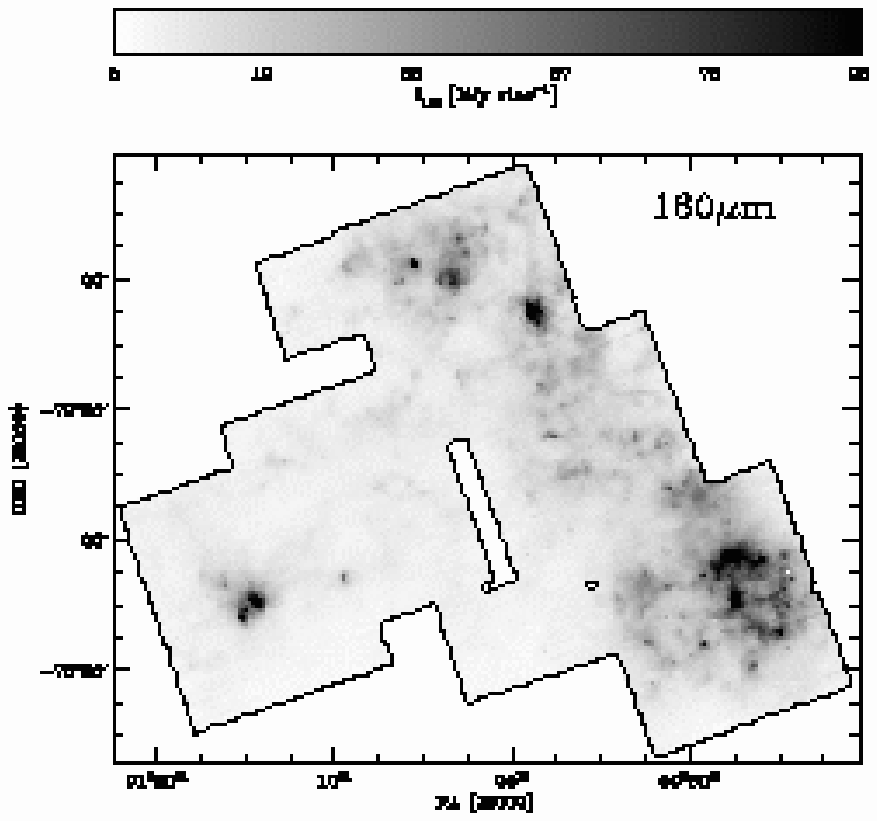}{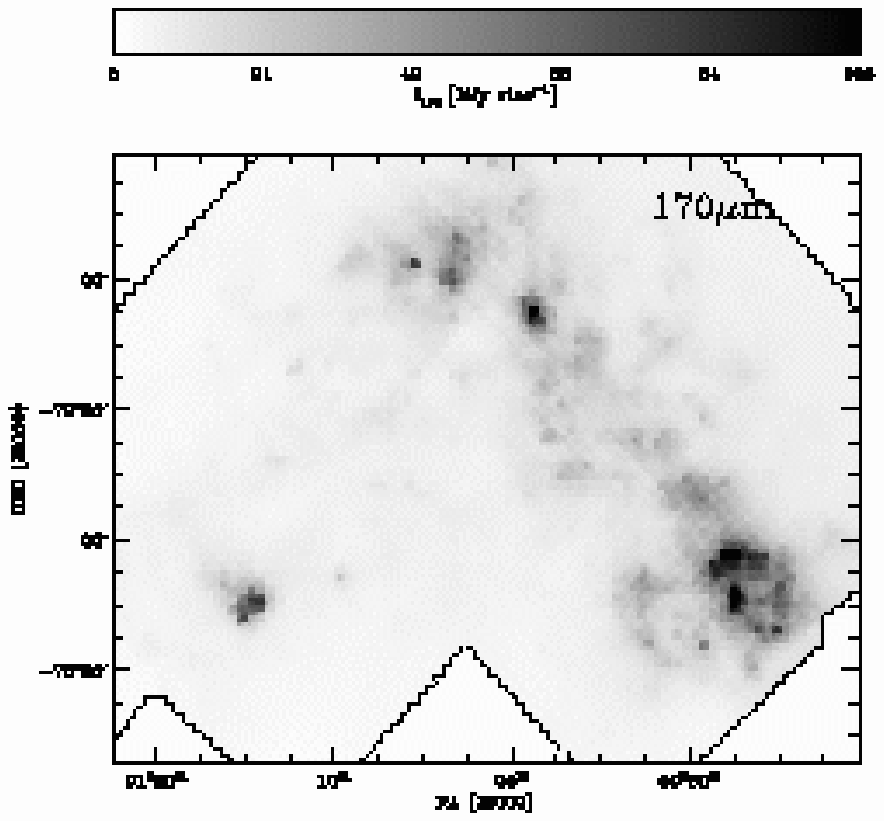}
 
\figcaption{\label{FIRMAPS} Maps of the SMC in six FIR bands arranged in order
  of increasing wavelength: (top left) the MIPS 24 $\mu$m map, (top right) the
  IRIS 60$\mu$m map, (middle left) the MIPS 70 $\mu$m map, (middle right) the
  IRIS 100 $\mu$m map, (bottom left) the MIPS 160 $\mu$m map, and (bottom
  right) the ISO 170 $\mu$m map \citep{BOT04}. The MIPS maps are shown at
  resolutions of $40\arcsec$ with the extent of the map shown by a black
  contour. The IRIS maps have $4\arcmin$ resolution. The ISO map has
  $1\arcmin.5$ resolution. All six maps have been corrected for zodiacal light
  and the Galactic foreground. The stretch is linear with the maximum for all
  maps selected to be 10 times the average intensity over a common area
  centered on the SW star-forming region.}

\end{center}
\end{figure*}

\begin{figure*}
\begin{center}
\epsscale{1.0}
\plottwo{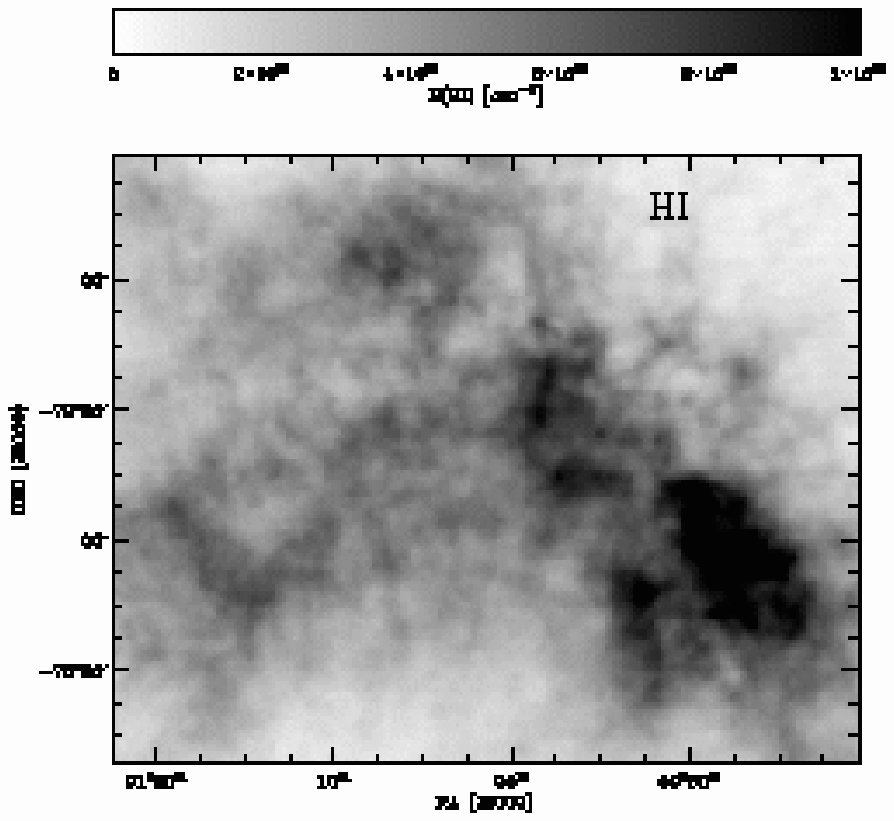}{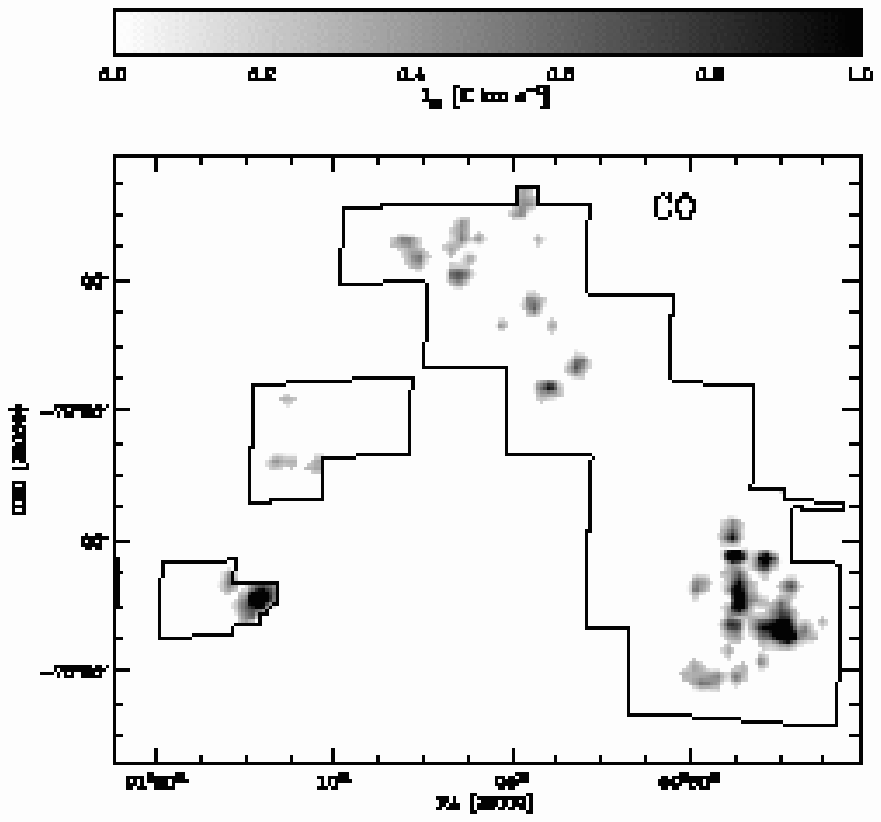}

\figcaption{\label{ISMMAPS} The ISM in the SMC: (left) the \hi\ at
$1\arcmin .5$ resolution \citep{STANIMIROVIC99}, and (right) \co\
intensity at $2\arcmin .6$ mapped by NANTEN \citep{MIZUNO01,MIZUNO06}. The
black contour shows the extent of the NANTEN survey.}

\end{center}
\end{figure*}

Figures \ref{FIRMAPS} and \ref{ISMMAPS} show FIR emission and neutral gas in
the SMC. Figure \ref{FIRMAPS} shows the MIPS and IRIS images after foreground
subtraction and processing. Figure \ref{ISMMAPS} shows \hi\ column density for
the SMC from \citet{STANIMIROVIC99} and the CO intensity map from the NANTEN
survey by \citet{MIZUNO01,MIZUNO06}. The CO map is masked using a threshold
contour of $0.3$ K km s$^{-1}$ (3$\sigma$), which is then expanded by
1$\arcmin$ in all directions.

\subsection{Dust Content of the SMC}
\label{DUSTMAPSECT}

Figure \ref{BESTMAP} shows dust surface density across the SMC. We
describe how we construct the map in Appendix
\ref{DUSTMAPAPP}. Briefly, we use the models by \citet{DALE02} to
derive the dust mass surface density across the SMC. These models
assume the dust along each line of sight to be heated by a power law
distribution of radiation fields; we estimate the power law index from
the 100-to-160 $\mu$m color. Appendix \ref{DUSTMAPAPP} also gives the
uncertainties associated with the dust map: a small $\lesssim 10\%$
statistical error, an additional 20\% systematic uncertainty from our
foreground subtraction, and a factor of 2 to 3 systematic uncertainty
associated with the choice of emissivity and model are the most
important. Over the region covered by the MIPS $160$ $\mu$m map, we
find a total dust mass of $M_{dust} =$\dustmass.

The SMC must have $M_{dust} \gtrsim 10^4$~M$_{\odot}$ of dust; this value
comes from assuming that a single population of grains is responsible for the
emission in the IRAS 60 and 100 $\mu$m bands. This is approximately the value
obtained by \citet{STANIMIROVIC00} and \citet{SCHWERING88}. In this case, the
emissivity will need to have a wavelength dependence of $\beta \sim 1$ in
order to match the emission observed at 140 $\mu$m and beyond.

Several estimates lead to an upper limit to the dust mass of $M_{dust}
\lesssim 10^6$ M$_{\odot}$. A Galactic dust-to-gas ratio, $DGR_{MW} \sim
10^{-2}$, applied to the mass of \hi\ in the SMC yields $M_{dust} \lesssim 4
\times 10^8~\mbox{M}_{\odot} \times 10^{-2} \approx 4 \times 10^6$. It is well
established that the $DGR$ in the SMC is lower than the Galactic one, so this
represents a firm upper limit. The heavy element abundance in the SMC is
$\approx 0.15$ that of the Galaxy \citep[measured from HII
regions,][]{DUFOUR84} and the total \hi\ content of the SMC is $4 \times 10^8$
M$_{\odot}$ \citep{STANIMIROVIC99}, so the total mass of heavy elements in the
ISM of the SMC is $\sim 1.2 \times 10^6$ M$_{\odot}$. Because the mass of dust
cannot exceed the mass of heavy elements, this represents another, similar,
upper limit to the dust content.

The observed sub-mm/millimeter SED also constrains the mass of dust. We
subtract the best fit ($\beta = 2$) single population model from the
integrated spectrum and fit a single population to the residuals at long
wavelengths. Assuming this emission comes from cold dust with a Galactic
emissivity ($\kappa_{250} = 8.5$ cm$^2$ g$^{-1}$ and $\beta = 2$), we find a
cold dust population with $T \approx 7$~K and $M_{dust} \approx
\colddustmass$. This is very close to the minimum equilibrium temperature
possible for interstellar grains \citep[$\sim 6$~K,][]{PURCELL69} so this
represents the largest dust mass permitted by the SMC SED.
\citet{STANIMIROVIC00} and \citet{WILKE04} both arrive at similar values for a
possible, but not necessary (see \S\ref{COLDDUSTSECT}), cold population, $\sim
10^6$ M$_{\odot}$ and $0.7 \times 10^6$ M$_{\odot}$, respectively
\citep[see][for the tradeoff between $T_{cold}$ and
$M_{cold}$]{STANIMIROVIC00}.

\begin{figure*}
\begin{center}
\epsscale{1.0} \plotone{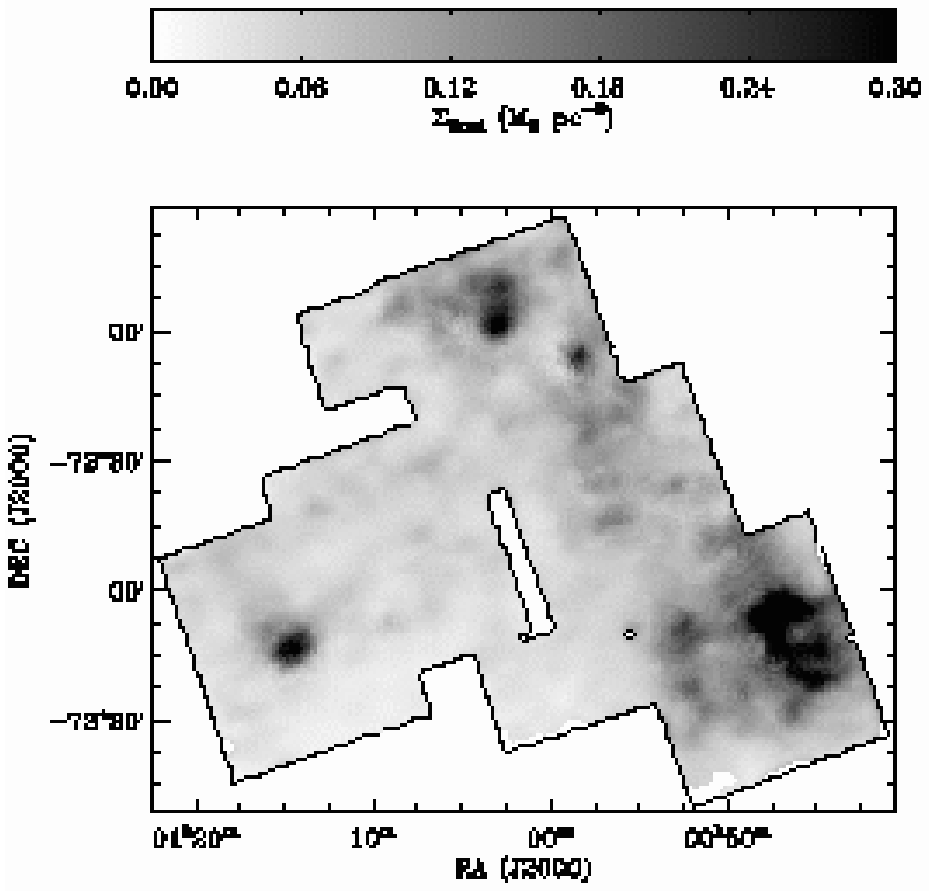}
 
\figcaption{\label{BESTMAP} The dust mass surface density map for the SMC at
  4$'$ resolution. The black line shows the extent of the MIPS 160 $\mu$m
  map.}

\end{center}
\end{figure*}

\subsection{The Integrated FIR SED of the SMC}
\label{INTSEDSECT}

Table \ref{FLUXTAB} gives the flux density integrated over the maps in the
wavebands used in this paper. We also give the range of literature values,
taken from \citet{WILKE04}, who present an excellent compilation of the FIR
SED of the SMC. Figure \ref{INTSED} shows log-log and log-linear plots of flux
density as a function of wavelength. In both Table \ref{FLUXTAB} and Figure
\ref{INTSED}, the area considered is the whole extent of the maps in Figure
\ref{FIRMAPS}, not just the smaller area covered by the MIPS maps. Because
this is an integrated SED over a large area, we are able to include points
from DIRBE and TopHat \citep[][]{AGUIRRE03}. We scale the latter by $0.9$ to
account for the difference between our area and the larger area studied by
TopHat; this is the mean ratio of DIRBE fluxes over the TopHat area to the
DIRBE fluxes over our own region. In deriving MIPS fluxes, we fill in the
areas not covered by the MIPS maps with DIRBE data interpolated to the MIPS
wavebands.  These filled in regions account for $\sim 20\%$ of the total flux
(the MIPS maps alone yield integrated fluxes of $310$, $10000$, and $15000$ Jy
at 24, 70, and 160 $\mu$m).

Figure \ref{INTSED} and Table \ref{FLUXTAB} show the photometric consistency
discussed in \S \ref{DATA} above. The data for each waveband agree within the
uncertainties. Figure \ref{INTSED} also shows the \citet{DALE02} models (solid
line) and best fit single populations with $\beta =$ 2, 1.5, and 1 (dotted,
dashed, and dash-dotted lines, respectively). All four models are derived from
only the integrated 100 and 160 $\mu$m data to isolate emission from big
grains. The best fit temperature for the big grains over the whole SMC is
$20.6$~K, somewhat higher than the $T=17.5$~K found for Galactic cirrus
\citep{BOULANGER96}. The best fit value for $\alpha$ over the whole SMC is
$2.4$, implying cirrus-like medium with a mean radiation field close to the
Solar Neighborhood value.

\subsubsection{The Long Wavelength Emission}
\label{COLDDUSTSECT}
The SMC emits more at submillimeter and millimeter wavelengths than one would
estimate from the FIR emission assuming Galactic dust ($\beta = 2$).  This may
typical for a dwarf irregular galaxy
\citep{BOLATTO00IC10,LISENFELD02,GALLIANO05}. This long wavelength emission
may indicate the presence of cold dust invisible in the FIR
\citep{STANIMIROVIC00,AGUIRRE03,WILKE04}; therefore, its interpretation is
relevant to this work. If the long wavelength emission is due only to cold
dust with a Galactic $\beta = 2$, then it could have $T_{cold} \sim 7$~K and
$M_{cold} \sim \colddustmass$, representing the dominant component of both
dust and heavy elements in the SMC. \citet{GALLIANO05} argued for this in
several galaxies similar to the SMC: that very cold (5 to 9~K) dust makes up
40 to 80\% of the dust mass.

\begin{figure}
\begin{center}
\epsscale{1.0}
\plotone{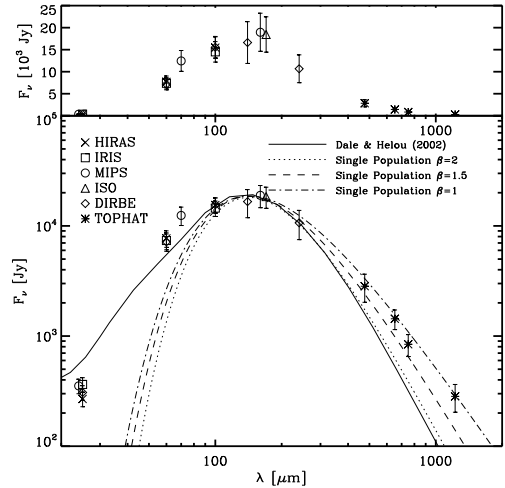}
 
\figcaption{\label{INTSED} The integrated SED of the SMC over the
region shown in Figure \ref{FIRMAPS} on linear (top) and logarithmic
(bottom) scales. We use DIRBE to fill in the (low intensity) regions
beyond the coverage of the MIPS maps. We scale the TopHat data
\citep{AGUIRRE03} by $0.9$ to match the area we consider. The
photometric consistency among the various bands is good. Nearby bands
agree within $1\sigma$ in all cases. The solid line shows the best fit
\citet{DALE02} model, derived from the 100 and 160 $\mu$m fluxes. The
three dotted lines show a single dust population with a temperature
inferred from the 100-to-160 $\mu$m color and emissivity power laws
with $\beta = 2,~1.5,$ and $1$.}

\end{center}
\end{figure}

However, Figure \ref{INTSED} shows that the long wavelength emission can also
be explained by an emissivity power law index shallower than $\beta = 2$ and
little or no cold dust. The best fit to the FIR adopting $\beta = 1.5$ yields
fluxes within 2$\sigma$ of every TopHat data point, though still
systematically low. The best fit $\beta$ from 100 $\mu$m to 1 mm with a single
population is $\beta \approx 1$ \citep[][find $\beta = 0.91$]{AGUIRRE03}.
Values of $\beta = 1$ and $1.5$ are within the range of plausible
astrophysical values, being appropriate for amorphous carbon and silicate
grains, respectively. Because the dust in the SMC is believed to be dominated
by silicate grains, a value of $\beta = 1.5$ may be more appropriate than
$\beta = 2.0$. Indeed, the best fit $\beta$ to the long wavelength values
$\kappa_{abs}$ found by \citet{WEINGARTNER01,LI01,LI02} for SMC grains beyond
$\sim 300 \mu$m is $1.65$. Further, \citet{ALTON04} argue that $\beta = 1.5$
describes emissivity determinations from the literature as well or better than
$\beta = 2.0$.

\citet{LISENFELD02, LISENFELD05} observed similar long wavelength emission in
NGC~1569, an irregular galaxy similar to the SMC in mass. They suggest that
the small, hot grains responsible for the 24 -- 70 $\mu$m emission may also
cause the long wavelength emission.  Using a modified version of the
\citet{DESERT90} models, they showed that an SED dominated by very small grain
(VSG) emission (with no PAHs) can fit the data as well as a model containing
cold big grains.  They argue that NGC~1569, a metal-poor galaxy with a strong
ISRF, is an unlikely locale to find a large amount of dust at unusually low
temperatures.  While such dust might hide in a few molecular regions across
the galaxy (we identify likely spots for this in the SMC below) there is a
dearth of places to hide the truly large amount of dust needed to account for
the long wavelength emission. Further, if such a reservoir of hidden does does
exist, it necessarily means that the metallicity of these galaxies has been
underestimated, in some cases substantially. On the other hand, in a
dynamically active environment with poor shielding, an enhancement of VSGs
relative BGs may be expected as enhanced erosion dust particle mantles turns
BGs into VSGs. These arguments may also apply to the SMC. However, the 25/60
$\mu$m and 60/160 $\mu$m colors in NGC~1569 ($\approx 0.13$ and $1.2$) are
higher than in the SMC ($\approx 0.05$ and $0.4$), more consistent with the
hypothesis of a large population of very hot dust. Simultaneous fitting of the
IRAC, MIPS, and IRAS data will be presented in a future paper.

Resolved sub-mm data is crucial to address the most compelling question raised
by \citet{LISENFELD02,LISENFELD05}: where, spatially, is the millimeter
emission coming from? The \citet{AGUIRRE03} results, with a beam of several
degrees, cannot address this question. At present, only one SMC molecular
cloud has published, resolved mapping at millimeter wavelengths.  This is
SMC~B1-1, mapped by SEST at 1.2~mm \citep{RUBIO04}. Towards this cloud, we
measure a background-subtracted flux of $F_{B1-1,160} \approx
1.5$~Jy\footnote[1]{In a $1\arcmin.5$ radius aperture about $\alpha_{1950} =
  0^h~43^m~42^s.4$, $\delta_{1950} = -73^{\circ} 35\arcmin 10\arcsec$
  \citep{RUBIO93}, with the background measured from a annulus between
  $5\arcmin$ and $6\arcmin$ to avoid contamination.}.  \citet{RUBIO04} find
$F_{B1-1,1200} = 50 \times 10^{-3}$~Jy. Therefore, $F_{160}/F_{1200} \approx
30$ for SMC~B1-1 compared to $F_{160}/F_{1200} \approx 65$ for the global SED
(see \S \ref{INTSED}). This GMC, described by \citet{RUBIO04} as cold and
quiescent, shows a millimeter excess relative to the whole SMC, consistent
with a cold dust temperature.  However, the excess is to small to explain the
the observed emission; extrapolating from SMC B1-1 --- $\sim 1\%$ of the CO
luminosity --- to the whole SMC yields a millimeter flux of only a few Jy.
\citet{AGUIRRE03} measure $280 \pm 80$ Jy at $\lambda = 1.2$mm.

We present three explanations for the long wavelength emission: a wealth of
cold dust, an abundance of hot small grains, and an emissivity law that
differs from the Galactic one. We cannot rule out any of these, but prefer the
latter, a modified emissivity, because it is simple and because the available
FIR/CO/mm data provide some evidence against the first and second
explanations.

\begin{deluxetable}{c l c}
\tabletypesize{\small}
\tablewidth{0pt}
\tablecolumns{5}
\tablecaption{\label{FLUXTAB} Integrated SED of the SMC}

\tablehead{\colhead{Waveband} & \colhead{Data Set} & \colhead{Flux Density} \\
$\mu$m & & Jy}

\startdata
24 & MIPS & $  350 \pm    50$ \\
\hline
25 & Literature\tablenotemark{a} & 256 -- 460 \\
25 & HIRAS & $ 270 \pm    40$ \\
25 & IRIS & $  360 \pm    60$ \\
25 & DIRBE & $  310 \pm    50$ \\
\hline
60 & Literature\tablenotemark{a} & 6689 -- 8450 \\
60 & HIRAS & $ 7700 \pm  1300$ \\
60 & IRIS & $7400 \pm  1300$ \\
60 & DIRBE &  $7200 \pm  1300$ \\
\hline
70 & MIPS & $12500 \pm  2400$ \\
\hline
100 & Literature\tablenotemark{a} & 12650 -- 16480 \\
100 & HIRAS & $15500 \pm  2200$ \\
100 & IRIS & $14500 \pm  2300$ \\
100 & DIRBE & $15600 \pm  2500$ \\
\hline
140 & Literature\tablenotemark{a} & 14000 -- 20030 \\
140 & DIRBE & $16600 \pm  4700$ \\
\hline
160 & MIPS & $19000 \pm  4300$ \\
\hline
170 & Literature\tablenotemark{a} & $15000 \pm 2300$ \\
170 & ISO & $18500 \pm  4000$ \\
\hline
240 & Literature\tablenotemark{a} & 9600 -- 12070 \\
240 & DIRBE & $10700 \pm  3200$ \\
\hline
476 & TopHat\tablenotemark{b} & $2800 \pm   800$ \\
652 & TopHat\tablenotemark{b} & $1400 \pm   300$ \\
750 & TopHat\tablenotemark{b} & $840 \pm   200$ \\
1224 & TopHat\tablenotemark{b} & $280 \pm   80$
\enddata

\tablenotetext{a}{The range of values from the literature compiled by
\citet{WILKE04} from \citet{SCHWERING88}, \citet{RICE88},
\cite{STANIMIROVIC00}, \citet{WILKE03}, and \citet{AGUIRRE03}.}

\tablenotetext{b}{Fluxes from TopHat scaled to our common area. The
total fluxes from \citet{AGUIRRE03} are 3200, 1620, 950, and 320 Jy.}

\end{deluxetable}

\subsection{Molecular Gas Map}

If dust and gas remain well mixed over the $\sim 400$~pc scales of our
calibration regions, and the properties of grains do not vary too dramatically
from the atomic to the molecular ISM, then the dust can be an optically thin
tracer of molecular gas. Dust measured from FIR emission offers several key
advantages over traditional molecular line tracers: unlike CO emission, dust
emission is optically thin; unlike molecules, dust is not expected to be
preferentially destroyed relative to H$_2$ in the outer parts of molecular
clouds \citep[e.g.][]{MALONEY88}; and the abundance of dust relative to gas
may be directly measured elsewhere in the galaxy via comparison with the \hi,
while the use of molecular lines as tracers of gas must be calibrated within
the GMCs themselves. In this section we use a method adapted from
\citet{ISRAEL97} and \citet{DAME01} to construct a map of the H$_2$ in the
SMC. We summarize the method, present the map, and then discuss the
uncertainties associated with the map.

\subsubsection{Method and Map}

To measure the H$_2$ mass surface density, $\Sigma_{\rm{H2}}$, along a line of
sight we first adopt a dust-to-hydrogen ratio, $D/H$. We apply $D/H$ to the
dust mass surface density to estimate the total (atomic plus molecular) mass
surface density along the line of sight, $\Sigma_{Total}$. This is

\begin{equation} 
\label{TOTGASEQN}
\Sigma_{Total} = \Sigma_{HI} + \Sigma_{\rm{H2}} =
\frac{\Sigma_{Dust}}{D/H}~.
\end{equation}

\noindent We remove the \hi\ column, $\Sigma_{HI}$, from each line of
sight to get a molecular gas column, $\Sigma_{\rm{H2}} = \Sigma_{Total} -
\Sigma_{HI}$. We measure $\Sigma_{HI}$ from the map by
\citet{STANIMIROVIC99}, which includes zero spacing data. So

\begin{equation} 
\label{H2EQN}
\Sigma_{\rm{H2}} = \frac{\Sigma_{Dust}}{D/H}~ - \Sigma_{HI}.
\end{equation}

\noindent We measure $D/H$ by comparing the mass surface density of dust
(Figure \ref{BESTMAP}) to the mass surface density of \hi\ over a series of
reference regions, attempting to simultaneously satisfy two conditions: 1)
that the reference regions is free of molecular hydrogen, and 2) that the
reference measurement take place near the region where we want to determine
$\Sigma_{\rm{H2}}$, in order to minimize the effect of spatial variations in
$D/H$.  We adopt the following method:

\begin{enumerate}
\item We identify regions likely to contain molecular gas from the CO map
  \citep[][]{MIZUNO01,MIZUNO06}. This is all lines of sight within $3
  \arcmin .6$ (65~pc) of $3\sigma$ CO emission. We derive $\Sigma_{\rm{H2}}$
  over each of these regions.

\item For each region, we create a reference region over which we measure
  $D/H$. This is all lines of sight more than $3\arcmin.6$ (65~pc) away from
  the CO but nearer than $10\arcmin.8$ (200~pc). The reference region is close
  enough that $D/H$ should apply to the H$_2$, but distant enough that
  contamination by H$_2$ is unlikely. We check this by stacking CO spectra
  across all of our reference regions: we measure $\bar{I}_{CO} \approx
  0.04$~K~km~s$^{-1}$, so there is only very diffuse CO emission in the
  reference region (compare to $\bar{I}_{CO} \approx 0.12$~K~km~s$^{-1}$ in
  the region outside of $3\sigma$ CO but within 65~pc).

\item Over each reference region, we measure the mean $D/H$ and derive
  $\Sigma_{\rm{H2}}$ using Equation \ref{H2EQN}.
\end{enumerate}

\begin{figure}
\begin{center}
\epsscale{1.0} \plotone{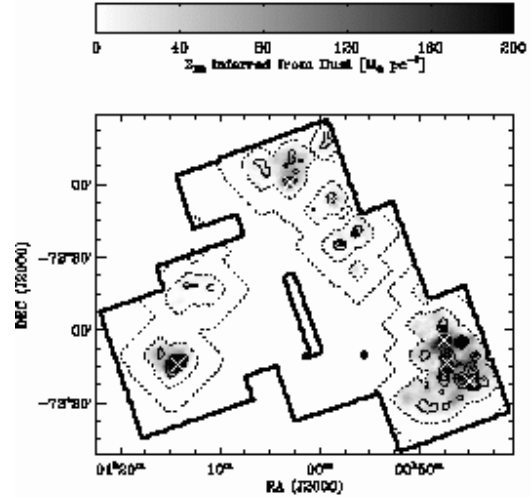}
 
\figcaption{\label{DUSTH2_METHOD} This figure illustrates the methods used to
  construct the map of $\Sigma_{\rm{H2}}$ shown in Figure \ref{DUSTH2}. Thin
  black contours show CO emission (at $0.3$, $0.6$, $0.9$, and $1.2$ K km
  s$^{-1}$), used to select regions over which we measure $\Sigma_{\rm{H2}}$.
  Dotted contours show the boundary of the reference regions (merged together)
  over which we measure local values of $D/H$. We calculate $\Sigma_{\rm{H2}}$
  only within the inner dotted boundary; the reference region is free of H$_2$
  by construction. Crosses indicate selected peaks from both the CO and H$_2$
  maps that will be used below. The grayscale shows the derived
  $\Sigma_{\rm{H2}}$.}

\end{center}
\end{figure}

\noindent Figure \ref{DUSTH2_METHOD} illustrates this method. We show a map of
the derived $\Sigma_{\rm{H2}}$, the CO emission (black contours), and an
outline of the set of all reference regions (dotted contours). Crosses mark
peaks in the H$_2$ and CO distributions (used below).

In order to match the resolution of the CO map, we derive $\Sigma_{\rm{H2}}$
from a high resolution version of our dust map. We measure the 100-to-160
$\mu$m ratio at the IRIS resolution, but we derive the dust surface density
(given the color) from the MIPS 160 $\mu$m map at the CO resolution. We assume
that the colors measured at $4\arcmin$ resolution apply to the map at
$2\arcmin.6$ resolution.  We performed the same calculation using the HIRAS
100 $\mu$m map, which has a $1\arcmin.7$ resolution. We derive a comparable
H$_2$ mass, but the map has a number of artifacts with scales comparable to
GMCs and therefore offers a check but no improvement.

Figure \ref{DUSTH2} shows the molecular gas mass surface density,
$\Sigma_{\rm{H2}}$, derived using this method. The white contours indicate the
extent of $3\sigma$ significant CO emission \citep[$0.3$ K km
s$^{-1}$][]{MIZUNO01,MIZUNO06} assuming the average \xco\ we derive below
(\intxco~\xcounits).  The H$_2$ map contains a total of \htwomass\ of
molecular hydrogen.  We estimate the plausible range of H$_2$ masses given our
systematic uncertainties as $\approx 1.5$ -- $4.0 \times 10^7$~M$_{\odot}$.

\subsubsection{Uncertainties and Systematics}

The statistical uncertainty in $\Sigma_{\rm{H2}}$ is small, only $\approx
6$~M$_{\odot}$~pc$^{-2}$. We estimate this from the measured uncertainties in
$\Sigma_{Dust}$ and $D/H$ and confirm it in two ways: 1) by introducing noise
at the $8\%$ level (see Appendix \ref{DUSTMAPAPP}) into $\Sigma_{Dust}$ and
deriving new $\Sigma_{\rm{H2}}$ maps for many realizations, and 2) by
computing $\Sigma_{\rm{H2}}$ for several regions known to be free of CO
emission. We also check the zero level in these empty regions, finding $-5$ to
$0$~M$_{\odot}$ pc$^{-2}$.

Systematic uncertainties in the FIR maps --- foreground, calibration, and
saturation --- will propagate to $\Sigma_{\rm{H2}}$. To evaluate these, we
offset the 100 $\mu$m and 160 $\mu$m maps by $\pm 1$ MJy ster$^{-1}$ one at a
time and rederive $\Sigma_{\rm{H2}}$. We perform similar tests to test the
effect of our calibration and saturation; we scaling first the whole $160$
$\mu$m map then just the bright regions by 1.25. Over all tests we find
$M_{\rm{H2}}$ in the range $1.4 \times 10^7$ to $3.9 \times 10^7$~M$_{\odot}$,
suggesting a $\sim 50\%$ uncertainty.

We avoid several systematic uncertainties by performing the determination of
the $D/H$ locally for each region. Nonetheless, several astrophysical
uncertainties may affect $\Sigma_{\rm{H2}}$: the choice of dust map, changes
in dust properties between the atomic and molecular ISM, and the possibility
of hidden cold dust.

The uncertainty in $\Sigma_{\rm{H2}}$ associated with our choice of a
particular dust map is less than 30\%.  We estimate this using test maps
adopting the various options described in Appendix \ref{DUSTMAPAPP} (i.e.
single population, only the 60 and 100 $\mu$m data, etc.). The total H$_2$
inferred from resulting maps spans $M_{\rm{H2}} = 2.1$ -- $3.2 \times
10^7$~M$_{\odot}$, with most values between $2.4$ and $3.2 \times
10^{7}$~M$_{\odot}$. Note that applying this method directly to the MIPS 160
$\mu$m intensity alone yields $M_{\rm{H2}} = 3.8 \times 10^7$~M$_{\odot}$.

Dust properties, especially the FIR opacity and $D/H$, may vary between the
atomic and molecular ISM, primarily due to growth of icy mantles on the
surfaces of dust grains in GMCs. These effects have the sense of decreasing
$\Sigma_{\rm{H2}}$ and a magnitude as high as a factor of 2 -- 3, though this
must be considered an upper limit because conditions in the SMC H$_2$ will be
less conducive to the growth of icy mantles than in Milky Way GMCs. See \S
\ref{DGRSECT} for more details.

If the possible population of very cold dust discussed in \S
\ref{COLDDUSTSECT} exists, our $\Sigma_{dust}$ map will be incomplete. If cold
dust is concentrated in GMCs, we will underestimate $\Sigma_{Dust}$ and thus
$\Sigma_{\rm{H2}}$ in these regions. In this case $\Sigma_{\rm{H2}}$ and
$\Sigma_{Dust}$ represent only lower limits. However, a large amount of cold
dust is not required to explain the SED (\S \ref{COLDDUSTSECT}) and indeed we
consider it unlikely.  In \S \ref{H2VSCOSECT} we find that quiescent gas with
a high CO-to-FIR ratio represents $\approx 30\%$ of the CO emission in the
SMC. If we allow that we miss all of H$_2$ associated with these clouds in our
map and assume that the average \xco\ we measure holds across the SMC, cold
dust seems will represent a $\sim 30\%$ correction to the overall H$_2$ mass.

Table \ref{H2UCTAB} lists the uncertainties discussed here and gives our
estimate of the maximum magnitude for each effect. \citet{ISRAEL97B} gives a
more thorough discussion of the uncertainties associated with this method.  He
concludes that most uncertainties have the effect that we underestimate the
true H$_2$ mass. We noted two exceptions, increased FIR opacity and $D/H$ in
GMCs, above.  Another is that we might mistakenly bookkeep warm ionized gas as
H$_2$. This requires very good resolution to investigate, so remains
uncertain.  However, studying the Eridanus Superbubble \citet{HEILES99} found
that IR not associated with \hi\ is mainly associated with H$_2$, not ionized
gas.

\begin{figure*}
\begin{center}
\epsscale{1.0} \plotone{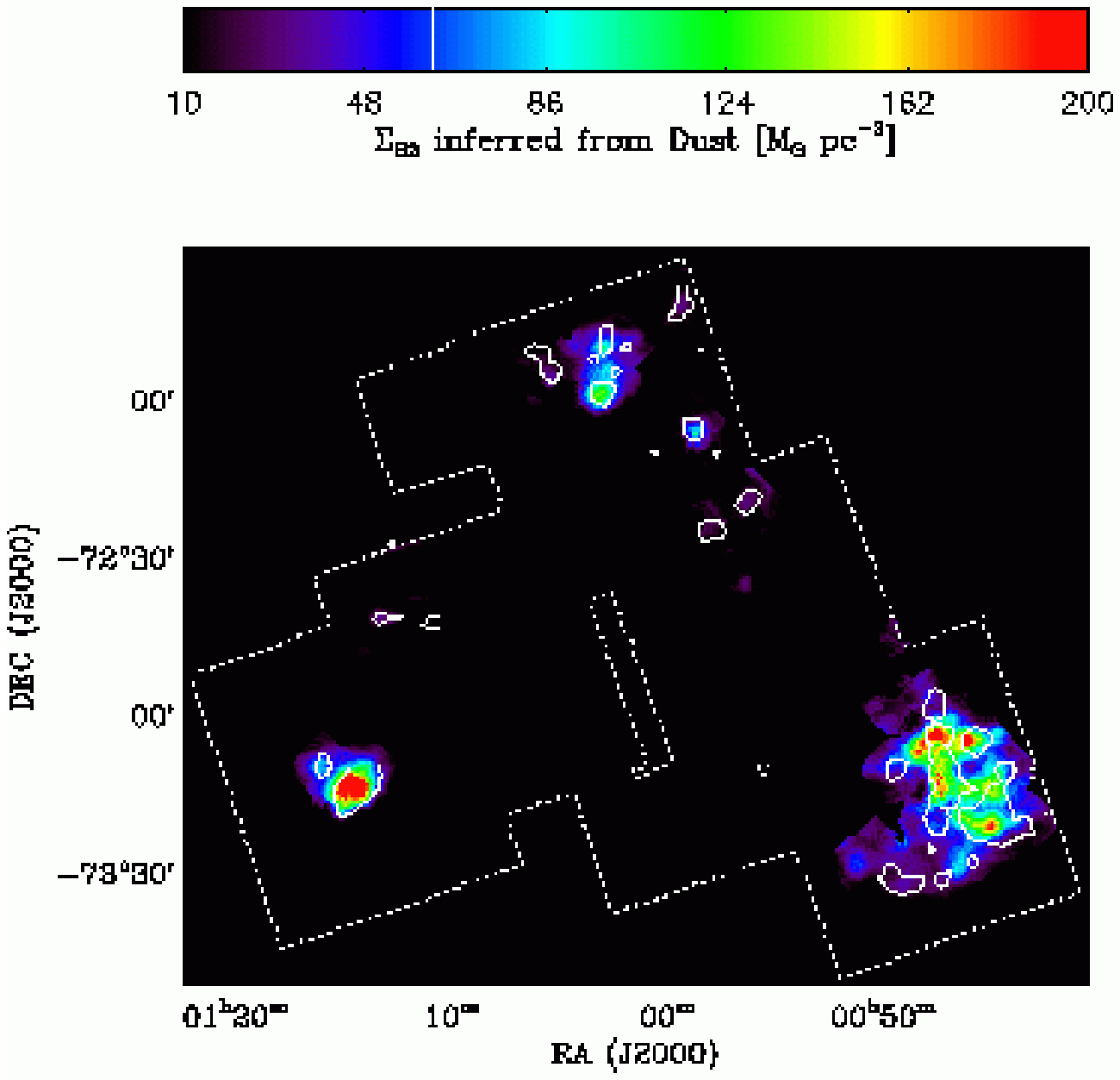}
 
\figcaption{\label{DUSTH2} The color image shows molecular gas mass
surface density in the SMC inferred from the dust, from 10 to 200
M$_{\odot}$ pc$^{-2}$. The 1$\sigma$ uncertainty in $\Sigma_{\rm{H2}}$
is $\sigma_{\Sigma H2} \approx 6$ M$_{\odot}$ pc$^{-2}$. The contours
show the extent of significant (3$\sigma$) CO emission, which is 0.3 K
km s$^{-1}$, or $\approx 62$~M$_{\odot}$ pc$^{-2}$ at $\xco =
\intxco$~\xcounits (see \S \ref{XCOSECT}). Both maps are at the
$2\arcmin.6$ resolution of the CO map. The dashed white contour shows
the extent of the MIPS 160 $\mu$m map.}

\end{center}
\end{figure*}

Several arguments suggest that our map of $\Sigma_{\rm{H2}}$ is not dominated
by these systematics. Most importantly, \citet{DAME01} found that H$_2$
estimated from FIR and CO emission agree well in the Milky Way, so the method
we apply to the SMC works in the Galaxy. The estimate of \xco\ that
\citet{DAME01} derives using this method is $1.8 \times 10^{20}$~\xcounits, in
excellent agreement with estimates based on gamma rays \citep{STRONG96}.
Further, \citet{SCHNEE06} studied the relation between FIR emission and dust
column for a simulated turbulent, externally heated cloud. They found that the
100/240~$\mu$m color (the best analog to our 100/160$\mu$m) does a good job of
matching the extinction for $A_V \lesssim 4$, though it may underpredict the
dust content above that value. Finally, Figure \ref{DUSTH2} shows good
agreement between our H$_2$ map and the CO map, particularly towards regions
of active star formation. In fact, toward these regions the H$_2$ is more
extended than CO, suggesting that when dust is warm enough to emit in the FIR,
the dust may trace H$_2$ better than CO.

Our estimate of the likely uncertainty in the $\Sigma_{\rm{H2}}$ map is
$50\%$. Systematic uncertainties in the data, dust modeling, and a modest
population of cold dust are all important but many of these effects have
offsetting senses.

\begin{deluxetable}{l c c}
\tabletypesize{\small}
\tablewidth{0pt}
\tablecolumns{5}
\tablecaption{\label{H2UCTAB} Uncertainties in the $\Sigma_{H2}$ Map}

\tablehead{\colhead{Description} & \colhead{Sense} &
  \colhead{Magnitude} }

\startdata
Data & & \\
\ldots Statistical\tablenotemark{a} & $\pm$ & 6 M$_{\odot}$ pc$^{-2}$ \\
\ldots Foreground & $\pm$ & $\left\{^{+10\%}_{-30\%}\right.$ \\
\ldots Calibration & $-$ & 30\% \\
\ldots Saturation & $+$ & 10\% \\
\ldots H$_2$ in Ref. Region & $+$ & $20$\% \\
\hline
Astrophysical & & \\
\ldots Model for Dust Map & $-$ & $30\%$ \\
\ldots Icy mantle growth in GMCs & divide & $(2$ -- $3)~f_{cold}$\tablenotemark{b} \\
\ldots Cold Dust (Best Guess) & $+$ & 30\% \\
\ldots Cold Dust (Maximum) & $+$ & $\Sigma_{\rm{H2}}$ is a lower limit

\enddata
\tablenotetext{a}{Per line of sight.}
\tablenotetext{b}{The fraction of molecular gas cold enough that accretion
  onto grains is significant, which we argue to be small.}
\end{deluxetable}

\section{Analysis}
\label{ANALYSIS}

\subsection{Dust and Atomic Gas}
\label{DGRSECT}

Over the region of the MIPS 160 $\mu$m map, the SMC has an \hi\ mass of
$M_{HI} = 1.8 \times 10^8$ M$_{\odot}$. The remainder of the $4 \times 10^8$
M$_{\odot}$ of \hi\ associated with the SMC \citep{STANIMIROVIC99} lies in an
extended component not covered by the MIPS map. With $M_{dust} = $\dustmass,
this implies $\log_{10} D/HI = \logdtog$, or 1-to-$\dtograt$ over the whole
SMC. If we include $M_{\rm{H2}} \approx 3 \times 10^7$, then $\log_{10} D/H =
\logdtogwhtwo$, or 1-to-$\dtogwhtworat$. Including helium, the total
dust-to-gas ratio is $\log_{10} DGR \approx \logdtogwhe$. This is lower than
the $DGR$ in the Milky Way by about the same factor as the metallicity
difference between the two systems with an uncertainty of a factor of $\sim 2$
(see Appendix \ref{DUSTMAPAPP}).  This result is consistent with measurements
by \citet{BOUCHET85}, who found a reddening per gas column $\sim 1/8$
Galactic, and by \citet{WILKE04}, who found a $DGR$ of 1-to-540.  There are
several predictions for how the $DGR$ varies with $Z$.  If we formulate $DGR
\propto Z^{a}$ then \citet{ISSA90} find $a = 1.2$, \citet{SCHMIDT93} find $a =
1.6$, \citet{LISENFELD98} find $a=1.9$, and \citet{DWEK98} find $a=1.3$. These
fits predict $\log_{10} DGR \approx -3.0, -3.0, -3.3, -3.5$. We cannot rule
out any of these values, but our derived $DGR$ leads us to prefer the
shallower \citet{ISSA90} and \citet{DWEK98} fits.

\begin{figure}
\begin{center}
\epsscale{1.0}
\plotone{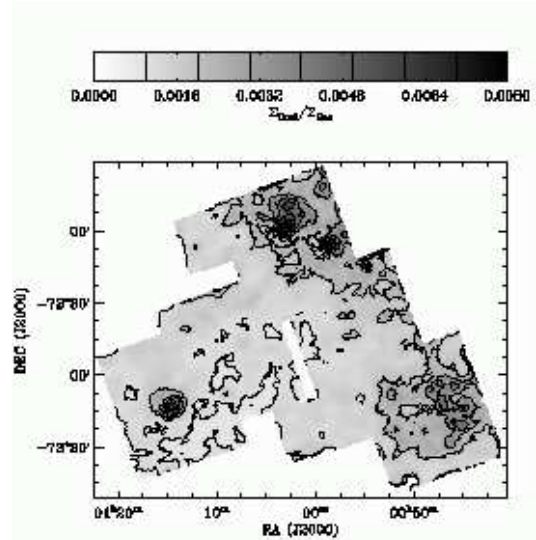}
 
\figcaption{\label{DUSTTOGASMAP} The ratio of dust mass surface density to
  atomic gas mass surface density at 4$\arcmin$ resolution. Star-forming
  regions show very high dust-to-gas ratios, probably because they contain
  molecular gas, which is not accounted for in this map. The star-forming Bar
  of the SMC shows higher $D/HI$ than the diffuse Wing.}

\end{center}
\end{figure}

\begin{figure}
\begin{center}
\epsscale{1.0}
\plotone{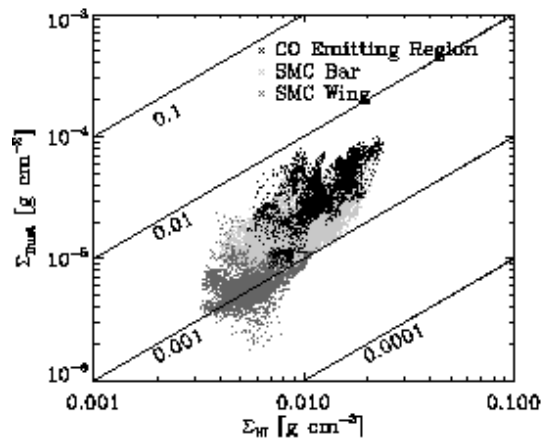}
 
\figcaption{\label{DUSTVSHI} Dust surface density as a function of \hi\
  surface density. Lines of sight towards CO emission, the SMC Bar, and the
  low intensity Wing are marked with black, light gray, and dark gray dots
  respectively. Lines of constant dust-to-gas ratio are labeled. The
  correlation between dust surface density and \hi\ is good across the SMC.
  The Bar shows a higher $D/HI$ than the Wing, and regions near CO emission
  show the highest dust-to-gas ratios, probably because they harbor
  significant amounts of molecular gas.}

\end{center}
\end{figure}

Figures \ref{DUSTTOGASMAP} and \ref{DUSTVSHI} compare dust and {\em atomic}
gas in our maps. Figure \ref{DUSTVSHI} shows that dust surface density and
atomic gas surface density are correlated across the SMC, with a rank
correlation coefficient $r_k\sim 0.76$ across all data.  All three figures
show that the $D/HI$ is a function of location in the SMC \citep[previously
noted by several authors and shown nicely in profile by][]{STANIMIROVIC00}.
Diffuse gas in the Wing --- distinguished by a simple intensity cut, $I_{160}
> 10$~MJy ster$^{-1}$ (which is arbitrary but does a good job) --- shows a low
mean $\log_{10} D/HI_{Wing} = \logdtogwing \pm 0.02$, while for the Bar
$\log_{10} D/HI_{Bar} = \logdtogbar \pm 0.01$, a factor of two higher. Lines
of sight towards regions with CO emission show higher $D/HI$ than nearby lines
of sight, the mean is $\log_{10} D/HI_{CO} = \logdtogco \pm 0.02$. It is
therefore possible to identify most of the SMC star-forming regions from the
map of $D/HI$, Figure \ref{DUSTTOGASMAP}.

Some of the variations in $D/HI$ may  be understood in terms of the life cycle
of  dust. \citet{DWEK98}  outlines  a picture  in  which dust  is depleted  in
supernova  shocks   and  accretes   new  material  within   molecular  clouds.
\citet{BOT04} argued that  supernovae shocks will destroy dust  more easily in
the diffuse gas found in the Wing. \citet{SEMBACH96} observed a similar effect
in the  Milky Way, that depletion  of some elements  decreases with increasing
distance from the plane of the Galaxy. This effect, combined with the distance
from the sources of new dust in the Bar, may explain the enhanced depletion in
the Wing.

The values of the $D/HI$ near CO emission must be due, at least in part, to
the presence of molecular gas; the presence of CO emission necessitates some
H$_2$ along the line of sight. If this is the dominant effect, then H$_2$
accounts for $\sim 30\%$ of the gas surface density in these regions, on
average.  The processing of dust inside GMCs may also contribute to the high
$\log_{10} D/HI_{CO}$: accretion of heavy elements onto grains may result in
an enhanced dust abundance and higher FIR opacities.

\citet{DWEK98} estimates that over the lifetime of a Milky Way GMC, the dust
mass in the cloud may increase by a factor of 2 -- 3 due to accretion of heavy
elements (see also references therein); so for Milky Way GMC in the middle of
its lifetime, we would expect an enhancement in the $DGR$ of $\sim 1.5$ --
$2$. However, an observational measure of the average $DGR$ enhancement
between the diffuse and molecular ISM is hard to come by. In the Milky Way,
\citet{RACHFORD02} found gas and dust to be well mixed along lines with $A_V
\gtrsim 1$. There is some evidence from absorption studies that the $DGR$
varies in the SMC with higher values toward lines of sight with more H$_2$
\citep{SOFIA06}, but the sample is too small to evaluate the significance of
this finding.

Some of the enhanced $D/HI_{CO}$ may also result from dust in GMCs having
larger FIR emission per unit mass at a given temperature (i.e., higher
$\kappa_{\rm{FIR}}$).  In the Milky Way, dust associated with dense, colder
gas is observed to have a higher emissivity than diffuse, warm gas
\citep{CAMBRESY01,STEPNIK03}.  Modeling the effect of icy mantles on grains in
protostellar cores, \citet{OSSENKOPF94} found an enhancement of a factor of
$2$--$3$ in the FIR opacity.

Both of these effects depend on conditions found at high $A_V$ which may be
more common in Milky Way GMCs than in the SMC. We shall see below that $A_V
\sim 1$--$2$ is typical for an SMC GMC, $\sim3$ times less on average than for
Milky Way GMCs. Therefore these conditions will not be as prevalent in SMC
H$_2$ as in the Milky Way. If $DGR$ enhancements and increased
$\kappa_{\rm{FIR}}$ are important effects, then the true $A_V$ towards SMC
GMCs is even lower. Therefore, we argue that the factors of $2$--$3$ discussed
here for both effects must be viewed as upper limits. In particular, the
\citet{OSSENKOPF94} calculations are for cores with densities $\gtrsim 10^5$
cm$^{-3}$, which will represent only a tiny fraction of the molecular gas in
the SMC. While all three effects contribute to the high value of $D/HI_{CO}$,
we consider it likely that presence of H$_2$ is primarily responsible for the
high $D/HI_{CO}$.

\subsection{CO and H$_2$ in the SMC}

\subsubsection{The CO-to-H$_2$ Conversion Factor}
\label{XCOSECT}

Our H$_2$ map contains \htwomass\ of molecular hydrogen. The CO luminosity in
the \citet{MIZUNO01,MIZUNO06} NANTEN map over the same area is $1.5 \times
10^5$ K km s$^{-1}$ pc$^{-2}$ (summing over the entire map without masking).
This implies a mean CO-to-H$_2$ conversion factor, $\xco$, of $\intxco$
\xcounits, which is $65$ times the Galactic value.  This conversion factor
varies systematically within the SMC. Regions with high CO intensity ($I_{CO}
> 0.3$~K~km~s$^{-1}$) have $\xco = 9 \pm 1 \times 10^{21}$ \xcounits, while
regions with diffuse CO emission ($I_{CO} < 0.3$~K~km~s$^{-1}$) have twice
this value, $\xco = 18 \pm 2 \times 10^{21}$~\xcounits.

If CO and H$_2$ are not coincident (see \S \ref{H2VSCOSECT}), then we must be
careful in defining exactly what is meant by the CO-to-H$_2$ conversion
factor. Indeed, much of the controversy in the literature is caused by
confusing large-spatial scale conversion factors, which may include H$_2$
without associated CO, with local conversion factors, which are limited to
regions with substantial CO emission. Therefore, we calculate a refined value
of \xco\ over the common volume shared by CO and H$_2$ for the molecular peaks
marked in Figure \ref{DUSTH2_METHOD}. We compare CO emission to H$_2$ in a
$1\arcmin$ diameter aperture, essentially measuring the beam pointed at the
center of the cloud. We then adjust the H$_2$ down by a factor of $\extfact$
to account for H$_2$ along the line of sight that lies outside of the CO
emitting region (we find in \S \ref{H2VSCOSECT} that the H$_2$ is more
extended than the CO by this factor). Table \ref{MPEAKTAB} shows the results.
We find a mean $\xco = \peakxco$~\xcounits. We derive a similar value, $\xco =
7 \pm 1 \times 10^{21}$~\xcounits, by considering lines of sight with bright
CO ($I_{CO} > 0.6$~K km s$^{-1}$) and scaling by the radius ratio
($\extfact$). Because it represents our best estimate of \xco\ over the volume
from which the CO emission originates, this is the value most directly
comparable to virial mass estimates.

Table \ref{XCOTAB} summarizes \xco\ from this paper and the literature. Most
of these values are from virial mass measurements, which obtain values of
\xco\ that are systematically higher than the Galactic value but dependent on
the scale of the structure studied \citep{RUBIO93, BOLATTO03}. Larger
structures yield higher values of \xco. Indeed, \citet{BOLATTO03}, studying
the N~83/N~84 region, found nearly Galactic values of \xco\ when considering
the smallest resolved structures in their data set (GMCs with radii of 10 to
20 pc), but at the scale of the entire complex found a conversion factor $\sim
10$ times the Galactic value.  At a common scale of $\sim 40$--$100$~pc, the
virial mass method yields $\xco = 0.9$--$3.5 \times 10^{21}$~\xcounits. At the
same scales, our FIR based H$_2$ map yields $\xco \approx \peakxco$~\xcounits,
$2$--$6$ times higher.

The discrepancy between M$_{H2}$ measured from the FIR and the virial mass may
be explained if magnetic fields may provide most of the support for SMC GMCs.
Magnetic support will not affect the turbulent line width, so in this case the
virial mass will underestimate the true mass of the GMC
\citep[e.g.][]{MCKEE92}. \citet{BOT06} compare H$_2$ masses estimated from
millimeter emission to CO-based virial masses for a set of SMC and Milky Way
GMCs. They find that in Milky Way clouds the virial mass usually exceeds the
millimeter-based mass, but that SMC clouds have higher millimeter masses than
virial masses. They attribute this difference to enhanced support by magnetic
fields in SMC clouds. If magnetic support does not play an important role, SMC
clouds may be short-lived as this measurement implies that the kinetic energy
is much smaller than the gravitational potential energy.

\citet{ISRAEL97} applied a method similar to ours to estimate M$_{\rm{H2}}$
towards several regions in the SMC with a limiting resolution of $15\arcmin$
($\sim270$~pc).  Given this resolution, his value of \xco = $12 \pm 2 \times
10^{21}$~\xcounits is best compared to our global value, $\intxco$~\xcounits,
and the two are identical within the uncertainties. Application of his method
to the newer data used in this paper yields $\xco = 9$ -- $13 \times 10^{21}$
\xcounits\ for lines of sight with CO emisqsion, depending on the method used
to define reference regions. The similarity between our results using the 100
and 160 $\mu$m bands and the \citet{ISRAEL97} results using the IRAS 60 and
100 $\mu$m bands shows that a color-corrected 100 $\mu$m map is enough to
capture the gross behavior of H$_2$ in a galaxy. This is also seen by the
success of \citet{DAME01} in matching Galactic FIR emission to the CO
distribution.  Longer wavelength (especially sub-mm) data may improve the
accuracy of column density estimates by $\sim 10\%$ and reduce the scatter
substantially \citep{SCHNEE06}, but is not necessary to obtain the basic
result. The improvements in the $\Sigma_{H2}$ map presented here are chiefly:
resolution capable of distinguishing Galactic GMCs ($\sim 50$~pc), allowing us
to study the relative structure of CO and H$_2$; and the use of local
reference regions to calibrate out variations in $D/H$ across the SMC --- the
low resolution of \citet{ISRAEL97} made a single $D/H$ appropriate for that
work.

\begin{deluxetable}{l c}
\tabletypesize{\small}
\tablewidth{0pt}
\tablecolumns{5}
\tablecaption{\label{XCOTAB} CO-to-H$_2$ Conversion Factor}

\tablehead{\colhead{Source} & \colhead{$\xco$} \\
& \xcounits }

\startdata
Whole SMC\tablenotemark{a} & $\intxco$ \\
LOS $I_{CO} > 0.3$ K km s$^{-1}$\tablenotemark{a} & $9 \pm 1 \times 10^{21}$ \\
LOS  $I_{CO} < 0.3$ K km s$^{-1}$\tablenotemark{a} & $18 \pm 2 \times 10^{21}$ \\
Molecular Peaks\tablenotemark{a,b} & $\peakxco$ \\
\hline
\citet{MIZUNO01}\tablenotemark{c} & $1$--$5 \times 10^{21}$ \\
\citet{BLITZ06}\tablenotemark{c} & $0.9$--$1.5 \times 10^{21}$ \\
\citet{RUBIO93}\tablenotemark{c} & $9.0 \times 10^{20}~\left(R/10~\mbox{pc}\right)^{0.7}$ \\
\citet{BOLATTO03}\tablenotemark{c} & \\
... Individual GMCs & $1.8$--$7.8 \times 10^{20}$ \\
... N83/N84 Complex & $2.2 \times 10^{21}$ \\
\hline
\citet{ISRAEL97}\tablenotemark{a,d} & $1.2 \times 10^{22}$ \\
\citet{RUBIO04}\tablenotemark{e} & $1.3$ -- $3.6 \times 10^{22}$ 
\enddata

\tablenotetext{a}{Using FIR to trace H$_2$.}

\tablenotetext{b}{Marked in Figure \ref{DUSTH2_METHOD}. Adjusted by
$\extfact$ to account for extended H$_2$ beyond CO along the line of
sight. Most appropriate for comparison with the virial mass.}

\tablenotetext{c}{Using virial mass method.}

\tablenotetext{d}{Using IRAS 60 and 100 $\mu$m bands.}

\tablenotetext{e}{Submillimeter continuum observations.}

\end{deluxetable}

\subsubsection{Comparison Between H$_2$ and CO}
\label{H2VSCOSECT}

Figure \ref{DUSTH2} shows that CO predicts the presence of H$_2$ almost
perfectly. However, the reverse is not true; we derive substantial H$_2$
surface densities beyond the $3\sigma$ CO emission (white contours). Thus,
H$_2$ appears more extended than CO emission in SMC star-forming regions.
However, an alternative explanation is hard to rule out: because \xco\ is very
high, the surface density sensitivity of the CO map may in fact be quite poor.
In this case, there may be low intensity CO below the noise level of the map
and the differences in the area are an artifact of sensitivity. In this
section we show that these differences are more than a sensitivity effect;
there are substantial variations in \xco\ across the SMC.

\begin{figure*}
\begin{center}
\epsscale{1.0}
\plotone{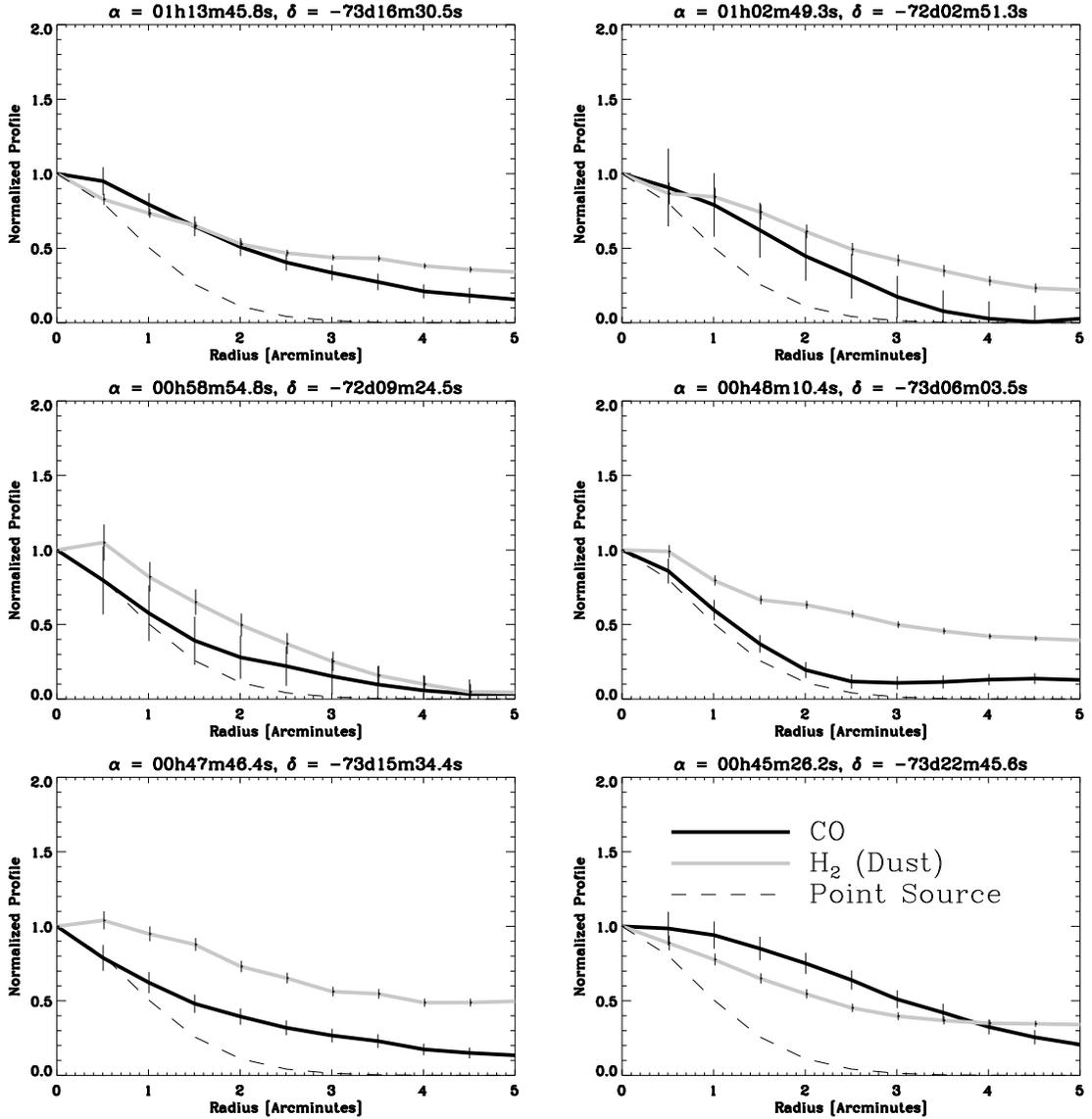}

\figcaption{\label{COPROFS} Normalized CO intensity (black) and H$_2$
surface density (gray) in concentric rings around selected maxima. All
profiles are normalized to the central value and then averaged in
concentric rings. The dashed lines show the profile of a point
source. The H$_2$ profile derived from the FIR and \hi\ is more
extended than the CO emission in 4 of 6 cases; in the other two cases
the profiles are comparable. The position of each maximum is indicated
above the panel and the maxima are noted with green crosses in Figure
\ref{DUSTH2}.}

\end{center}
\end{figure*}

Figure \ref{COPROFS} shows the distribution of H$_2$ and CO about several
molecular gas peaks. These peaks are indicated in Figure \ref{DUSTH2_METHOD}
and summarized in Table \ref{MPEAKTAB}. We normalize each profile by its peak
value, so Figure \ref{COPROFS} contains no information about the relative
amplitude of the two distributions, i.e. \xco. Instead, Figure \ref{COPROFS}
shows the relative structure in the two maps. For four of the six peaks, H$_2$
is more extended than CO; around one peak CO is more extended than H$_2$; and
around one peak, the N~83/N~84 region in the east, CO and H$_2$ have similar
structure.

We stack the normalized profiles and measuring the half-width at half maximum
(HWHM) for the average H$_2$ and CO profiles.  The stacked H$_2$ has
HWHM$\sim41$~pc while the stacked CO has HWHM$\sim30$~pc. Integrating the
average H$_2$ profile and the average CO profile we find $\int
\Sigma_{\rm{H2}} / \int \Sigma_{CO} = \extfact$. Thus, along an average line
of sight towards a molecular peak, $\sim 30\%$ of the H$_2$ lies outside of
the region with CO emission --- a number we used to compute the ``common
volume'' \xco\ estimate above.

\begin{figure}
\begin{center}
\epsscale{1.0} \plotone{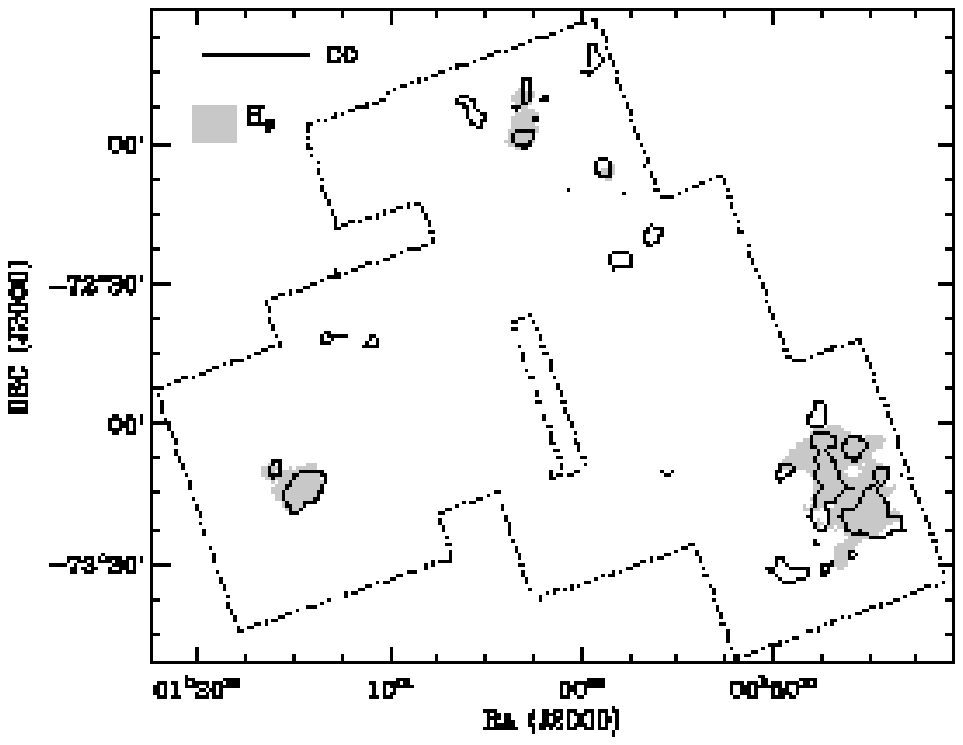}

\figcaption{\label{HISIGREGIONS} The extent of CO and H$_2$ from the
FIR at a common significance assuming our global CO-to-H$_2$
conversion factor, $\xco = \intxco$~\xcounits. Both maps
are clipped at the $3\sigma$ for the CO, or $\Sigma_{\rm{H2}} \approx
60$~M$_{\odot}$~pc$^{-2}$. In regions of active star formation, the
H$_2$ shows a greater extent than the CO, but away from these regions,
CO emission may indicate the presence of cold(er) dust.}

\end{center}
\end{figure}

Figure \ref{HISIGREGIONS} shows another view of the relative distribution of
H$_2$ and CO. We plot regions with $\Sigma_{\rm{H2}} >
60$~M$_{\odot}$~pc$^{-2}$ in gray and indicate the extent of significant CO
emission ($3\sigma = 0.3$~K~km~s$^{-1}$) with black contours. These two
thresholds are matched for $\xco = \intxco$~\xcounits, our integrated
conversion factor. Note that this is $\sim 2$ times higher than our estimate
of \xco\ over the shared volume. Therefore, the gray area in Figure
\ref{HISIGREGIONS} shows a very conservative estimate of where would expect to
find CO emission based on our FIR-derived $\Sigma_{\rm{H2}}$. Despite this
conservative cut, toward regions of active star formation --- at the north and
south of the Bar and in the eastern N~83 region --- H$_2$ extends beyond the
CO. In these regions, the dust is warm and Figures \ref{HISIGREGIONS} and
\ref{DUSTH2} give us confidence that our FIR-based H$_2$ map traces
$\Sigma_{\rm{H2}}$ better than CO.

Thus we present three measurements that suggest the H$_2$ to be more extended
than the CO: 1) the profiles shown in Figure \ref{COPROFS}; 2) the large
extent of CO relative to H$_2$ at a common, conservative threshold in Figure
\ref{HISIGREGIONS}; and 3) most compellingly, the measurement of different
values of \xco\ averaging over low and high $I_{CO}$ regions --- where $I_{CO}
> 0.3$~K~km~s$^{-1}$ we find $\xco = 9 \pm 1 \times 10^{21}$ \xcounits\ and
where $I_{CO} < 0.3$~K~km~s$^{-1}$ we find $\xco = 18 \pm 2 \times
10^{21}$~\xcounits. None of these measurements depend critically on the
amplitude of the H$_2$ map, i.e. $M_{\rm{H2}}$, or the sensitivity of the CO
map, but only on the relative structure of the H$_2$ and CO maps. We interpret
this increase in H$_2$ relative to CO towards the outer parts of SMC clouds as
evidence of the selective photodissociation of CO in these regions
\citet{MALONEY88}.

Away from the star-forming regions, there is more CO relative to FIR-derived
H$_2$. Several small CO clouds --- in the Wing, the central Bar, and outside
the northern and southern star-forming regions --- may be good candidates to
harbor populations of cold dust. In these regions, we still find significant
H$_2$ emission, --- $90\%$ of lines of sight with $I_{CO} > 0.3$~K~km~s$^{-1}$
also have $\Sigma_{\rm{H2}} > 20$~M$_{\odot}$~pc$^{-2}$ (3$\sigma$) and the
remainder all have $\Sigma_{\rm{H2}}$ at slightly lower significant.  However,
towards these regions we may underestimate the total $\Sigma_{\rm{H2}}$
because there is a population of cold dust that is incompletely traced by our
FIR maps. 

How much molecular gas is represented by these quiescent clouds? The region
with $\Sigma_{\rm{H2}} > 60$~M$_{\odot}$~pc$^{-2}$ contains both 70\% of the
significant CO emission and 70\% of the significant H$_2$ surface densities.
While the FIR-based $\Sigma_{\rm{H2}}$ may fail in these regions, they will be
subject to less intense radiation fields and should therefore suffer less
selective photodissociation of CO. Therefore, based on the fraction of the
total CO luminosity coming from such clouds, we estimate the fraction of the
SMC molecular phase in cold, quiescent clouds to be $\sim 30\%$.

\begin{deluxetable*}{l l c c c c c l}
\tabletypesize{\small}
\tablewidth{0pt}
\tablecolumns{5}
\tablecaption{\label{MPEAKTAB} Properties Of Several Molecular Peaks}

\tablehead{\colhead{$\alpha$} & \colhead{$\delta$} &
  \colhead{$\Sigma_{\mbox{HI}}$} & \colhead{$\Sigma_{\mbox{H2}}$} &
  \colhead{$\xco$\tablenotemark{a}} & HWHM(H$_2$) & HWHM(CO) & $\log_{10} D/H$ \\~~(J2000) & (J2000) & 
  (M$_{\odot}$ pc$^{-2}$) &  (M$_{\odot}$ pc$^{-2}$) & $\frac{10^{21} \mbox{cm$^{-2}$}}{\mbox{K km s$^{-1}$}}$ &
  (pc) & (pc) & (Reference) }

\startdata
01h13m45.8s & -73d16m30s & 58 & 261 & 7.1 & 40 & 36 & -2.93 \\
01h02m49.3s & -72d02m51s & 49 & 133 & 10 & 44 & 33 & -2.70 \\
00h58m54.8s & -72d09m24s & 36 & 81 & 5.4 & 35 & 22 & -2.64 \\
00h48m10.4s & -73d06m03s & 106 & 230 & 5.6 & 53 & 22 & -2.82 \\
00h47m46.4s & -73d15m34s & 83 & 162 & 4.0  & 69 & 26 & -2.82 \\
00h45m26.2s & -73d22m45s & 70 & 199 & 6.5  & 40 & 54 & -2.82
\enddata
\tablenotetext{a}{Applying a correction of $1/1.3$ to account for H$_2$ beyond
  the CO along the line of sight (see text).}

\end{deluxetable*}

\subsubsection{Surface Density and Extinction}

If photoionization by far-UV photons regulates the structure of molecular
clouds, then GMCs in low metallicity systems may be expected to have higher
surface densities than Milky Way GMCs in order to achieve the same mean
extinction as their Galactic counterparts \citep[average ${A_{V}} \approx 4$
-- $8$ over a Milky Way GMC,][]{MCKEE89,SOLOMON87}. \citet{PAK98} estimated
$N(H)\gtrsim 10^{23}$ cm$^{-2}$ ($\Sigma_{H}\gtrsim800$ M$_{\odot}$ pc$^{-2}$)
towards SMC star-forming regions. Using our dust and gas maps, we calculate
the mean surface density and extinction towards the molecular peaks in the
SMC.

We find surface densities similar to those of spiral galaxy GMCs rather than
the values of $\gtrsim 1000$ M$_{\odot}$ pc$^{-2}$ expected from
\citet{MCKEE89} and \citet{PAK98}. The molecular gas peaks indicated in Figure
\ref{DUSTH2_METHOD} have surface densities $\Sigma_{\rm{H2}} = 90$ -- $310$
with a mean $\Sigma_{\rm{H2}} = 180 \pm 30$ M$_{\odot}$ pc$^{-2}$. This
similar to the mean for Milky Way GMCs, $\Sigma_{\rm{H2}} \approx 170$
M$_{\odot}$ pc$^{-2}$ \citep{SOLOMON87}. If CO-dark H$_2$ makes up 10 -- 50\%
of the molecular gas in the Milky Way \citep{GRENIER05}, SMC clouds may even
have $\Sigma_{H2}$ slightly lower than Galactic clouds. Along the same lines
of sight, the \hi\ surface density varies from $\Sigma_{HI} = 35$ -- $105$
M$_{\odot}$ pc$^{-2}$ with a mean of $\Sigma_{HI} \approx 70 \pm 10$
M$_{\odot}$ pc$^{-2}$. Over the entire region with 3$\sigma$ CO emission --- a
reasonable estimate of the star-forming area --- we find a mean
$\Sigma_{\rm{H2}} = 90 \pm 10$ M$_{\odot}$ pc$^{-2}$ and a mean $\Sigma_{HI} =
65 \pm 2$ M$_{\odot}$ pc$^{-2}$.

Nearly Galactic surface densities imply that the mean extinction through SMC
GMCs is much lower than in Galactic GMCs.  \citet[][]{BOUCHET85} found $R_V =
A_V / E(B-V) = 2.7 \pm 0.2$ for the SMC. Using this value and $N_H / E(B-V) =
5.8 \times 10^{21}$ cm$^{-2}$ mag$^{-1}$ \citep[][]{BOHLIN78}, scaled down by
the $D/H$ relative to Galactic (taken here to be $\approx 0.01$), we calculate
$A_V$:

\begin{equation}
\label{HITODUSTEQN}
A_V \approx ~\frac{N(HI) + 2~N(H_2)}{2.1 \times
  10^{21}~\mbox{cm}^{-2}}~\frac{D/H_{SMC}}{D/H_{MW}}~\mbox{.}
\end{equation}

\noindent We measure $D/H_{SMC} / D/H_{MW}$ to be $\approx 7$ (\S
\ref{DUSTMAPSECT}). Using Equation \ref{HITODUSTEQN}, we find $A_V \approx 1.8
\pm 0.6$ mag towards the molecular peaks and $A_V \approx 1.1 \pm 0.5$ mag
over the whole region with CO emission. Apply the method of \citet{SCHLEGEL98}
to our dust map yields consistent extinctions, $A_V \approx 1.6 \pm 0.3$
toward the peaks and $A_V \approx 1.0 \pm 0.4$ over all CO emission.

Our H$_2$ map has $46$~pc resolution, close to the average diameter of a Milky
Way GMC \citep{SOLOMON87}. If the star-forming parts of SMC GMCs have the same
diameters as Milky Way clouds, then our data are not consistent with the high
surface densities predicted by \citet{MCKEE89} for low metallicity GMCs.
However, the region of CO emission, rather than the more extended H$_2$
region, may be the relevant comparison for the \citet{MCKEE89} prediction. We
only expect star formation to occur within the CO-emitting region, and the
$\bar{A}_V = 4$ -- $8$ calculated by \citet{MCKEE89} is for GMCs defined by CO
emission. In this case, resolution effects (``beam dilution'') may cause us to
underestimate the surface density associated with the CO peaks. If the radius
of a typical SMC GMC is $\sim 15$~pc, for example, then the associated surface
density of H$_2$ may be $\sim 1600$~M$_{\odot}$~pc$^{-2}$, consistent with the
\citet{MCKEE92} prediction. Many structures in the \citet{MIZUNO06} CO map are
unresolved or only marginally resolved, and other studies have indicated that
CO clouds in both LMC and SMC may be smaller than in the Milky Way, with CO
confined to small regions and little or no diffuse component
\citep[e.g.][]{LEQUEUX94,RUBIO99,MIZUNO01}.

Thus, we find evidence for differences between SMC and Milky Way GMCs, but
cannot distinguish the sense of them because of limited resolution. SMC GMCs
either have lower average extinction through the clouds or are substantially
($\sim 3$ times) smaller on average than Milky Way GMCs. A second high
resolution band is needed to test which of these conclusions holds by
measuring the surface densities of SMC GMCs at a resolution of $\approx 10$~pc
(that of the MIPS 160 $\mu$m map).

\subsubsection{Pressure and Molecular to Atomic Gas Ratio}

In spiral galaxies there is a good correlation between the ratio of
molecular gas mass to atomic gas mass and the hydrostatic midplane
pressure \citep{BLITZ04}. Dust is integral to H$_2$ formation and may
impede H$_2$ destruction (though self-shielding of H$_2$ may also have
this effect). Therefore we expect that, given the low dust-to-gas
ratio in the SMC, the ratio of molecular gas to atomic gas for a given
pressure (gas density) will be lower in the SMC than in spiral
galaxies.

\citet{BLITZ04} derive that the hydrostatic midplane pressure for a
thin disk of gas in a larger disk of stars is

\begin{equation}
\label{PRESSEQN}
P_{h} =  0.84 (G \Sigma_*)^{0.5}\Sigma_g \frac {v_g} {(h_*)^{0.5}}
\end{equation}

\noindent where $\Sigma_{g}$ is the total surface density of the gas,
$\Sigma_{*}$ is the surface density of stars, $v_g$ is the velocity
dispersion of the gas, and $h_*$ is the scale height of the stellar
disk. The applicability of this formula in the SMC is questionable:
the three dimensional distribution of the stellar potential is likely
not well described as a disk. Indeed, \citet{CROWL01} find a 1$\sigma$
dispersion of 6 -- 12 kpc along the line of sight through the
SMC. They suggest that the galaxy may be a triaxial spheroid with a
1:2:4 ratio of sizes along the right ascension, declination, and
line-of-sight directions. Also, although there is usually only a
single molecular complex along a given line of sight, there are often
several \hi\ components. Even if Equation \ref{PRESSEQN} applies, the
inclination, stellar scale height, and the mass-to-light ratio of the
stellar and gas disks are not known to better than a factor of two.

Emphasizing the approximate nature of the calculation, we derive the
pressure and the ratio of molecular to atomic gas at $\sim 0^\circ.7$
resolution ($\sim750$~pc). The mean velocity dispersion of \hi\ along
the line of sight is $v_g\approx20$ km s$^{-1}$
\citep{STANIMIROVIC99}. We adopt $h_* = 6$~kpc \citep[the low end of
the][range]{CROWL01} and no inclination. We use the DIRBE $K$-band map
and a mass-to-light ratio $M/L_K = 0.5~M_{\odot}/L_{K,\odot}$
\citep[e.g.][]{SIMON05} to calculate $\Sigma_*$, and the FIR-based
H$_2$ map combined with the \hi\ map to calculation the gas surface
density and the ratio of molecular to atomic gas. 

\begin{figure}
\begin{center}
\epsscale{1.0}
\plotone{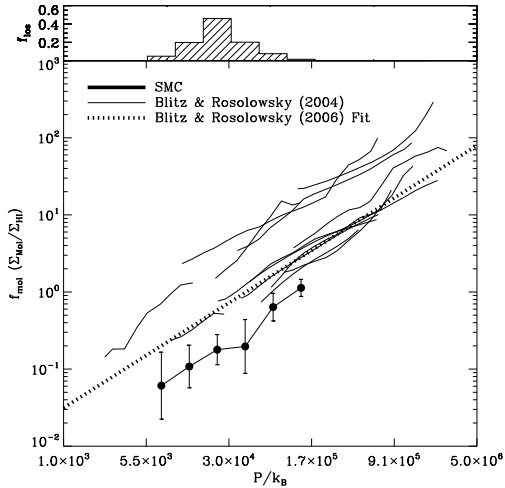}

\figcaption{\label{PRESS} The molecular to atomic gas ratio,
$f_{\rm{mol}}$, as a function of hydrostatic midplane pressure,
conducted at $0^{\circ}.7$ resolution. We use the formula of
\citet{BLITZ04} with $\sigma_v = 20$~km~s$^{-1}$ and $h_* = 6$~kpc and
also plot data for spiral galaxies measured at comparable spatial
resolution \citep{BLITZ06B}. The histogram above the plot shows the
fraction of lines of sight in each pressure bin for the SMC. The low
H$_2$ to \hi\ ratio in the SMC is a result of the relatively low
values of $P_{h}$ and an additional factor of $\sim 3$ deficit that
may be the result of a low dust-to-gas ratio.}

\end{center}
\end{figure}

Figure \ref{PRESS} shows $f_{mol}$ as function of $P_{h}$ for the SMC and a
sample of spiral galaxies studied by \citet{BLITZ06B} with similar spatial
resolution. Most lines of sight for the SMC have $P_{h} / k_B = 10^4$ --
$10^5$~K~cm$^{-3}$, with almost half of the data in the bin centered on
$P_{h}/k_B = 2.4 \times 10^4$~K~cm$^{-3}$. In this bin, the best fit
\citet{BLITZ06B} relation predicts $f_{mol} \approx 0.6$ for large spirals.
For reference, this $P_{h}$ and $f_{mol}$ are similar to those found at the
Solar Circle in the Milky Way, a region that is dominated by \hi\ by about
2-to-1. In the SMC we measure $f_{mol} = 0.2$ in the bin centered on
$P_{h}/k_B = 2.4 \times 10^4$~K~cm$^{-3}$, a factor of $3$ lower than the
prediction/Milky Way value. Over the whole SMC, we find $f_{mol}$ a factor of
$2$ -- $3$ lower than predicted.

Thus, Figure \ref{PRESS} suggests that the low ratio of H$_2$ to \hi\ in the
SMC is due to a combination of low gas densities and environmental effects.
Because the median P$_{h}$ in the SMC is lower than in the inner disks of
spiral galaxies, $f_{mol}$ is low. However, $f_{mol}$ is lower than found in
spiral galaxies by an additional factor of $2$ -- $3$. This may be the result
of the low $DGR$. We suggest the following explanation: Because the formation
rate of H$_2$ is limited by the rate at which H atoms stick to the surface of
dust grains \citep{HOLLENBACH71}, H$_2$ will form more slowly in the dust-poor
SMC. This lower formation rate means that given the same conditions ---
density, shielding from dissociating radiation --- the equilibrium H$_2$/\hi\
ratio should be lower for a low $DGR$ galaxy. Alternatively, SMC clouds may be
out of equilibrium: for a low $DGR$, an \hi\ cloud will take longer to convert
to H$_2$, even under conditions where an overwhelmingly molecular cloud is the
equilibrium state \citep[][]{BELL06}. If the timescale to reach equilibrium is
long compared to the cloud lifetime, then these clouds may never reach
chemical equilibrium. In this case, we may expect to find a lower molecular
gas to atomic gas ratio in low metallicity galaxies like the SMC.

\section{Conclusions}
\label{CONCLUSIONS}

We present new FIR maps of the SMC at 24, 70, and 160 $\mu$m obtained by {\em
  Spitzer} as part of the S$^3$MC survey \citep{BOLATTO06}. These maps cover
the Bar and the Wing regions with $6\arcsec$, $18\arcsec$, and $40\arcsec$
resolution at 24, 70, and 160 $\mu$m respectively (1.8, 5.4, and 12 pc).
Combining these maps with data from the literature, we draw the following
conclusions:

\begin{enumerate}

\item We find a total $M_{dust} = $\dustmass, which is systematically
  uncertain by a factor of $2-3$ because of degeneracies in how to model the
  dust. Strong upper and lower limits to the dust mass in the SMC are $\sim 2
  \times 10^4$ and $10^6$ M$_{\odot}$, respectively. 

\item If the long wavelength part of the SMC SED comes from a population of
  very cold dust, then $M_{dust}$ is a lower limit.  However, based on the
  lone GMC with published, resolved millimeter observations it seems unlikely
  that GMCs harbor the requisite amount of cold dust to account for the long
  wavelength emission. There are plausible alternative explanations: an
  abundance of hot VSGS or that the emissivity of SMC grains at long
  wavelengths differs from the Galactic value. The latter explanation is
  simple and consistent with the data.

\item The dust-to-hydrogen ratio, $D/H$, over the whole region studied is
  $\log_{10} D/H = \logdtogwhtwo$, or 1-to-\dtogwhtworat, including H$_2$,
  implying a total $\log_{10} DGR \approx -3.0$. This is about
  $Z_{SMC}/Z_{MW}$ times the Milky Way $DGR$ and is consistent with
  \citet{BOUCHET85} and \citet{WILKE04}.

\item The dust-to-\hi\ ratio, $D/HI$, is a function of position in the SMC.
  In the Wing and the Bar, we find mean $\log_{10} D/HI_{Wing} = \logdtogwing
  \pm 0.02$ and $\log_{10} D/HI_{Bar} = \logdtogbar \pm 0.01$. $D/HI$ near
  regions with CO emission is higher still: $\log_{10} D/HI_{CO} = \logdtogco
  \pm 0.02$.  Some of this high value may be due to enhancements in the
  opacity and abundance of dust in molecular clouds, e.g. due to growth of icy
  mantles.  However, we argue that most of the $D/HI$ enhancement near CO
  emission is best explained by the presence of molecular hydrogen.

\item Following a method outlined by \citet{ISRAEL97} and \citet{DAME01}, we
  derived a map of $\Sigma_{\rm{H2}}$ in the SMC. To derive this map, we
  calibrate $D/H$ locally using a set of reference regions show in Figure
  \ref{DUSTH2_METHOD} and apply Equations \ref{TOTGASEQN} and \ref{H2EQN}.  We
  find $M_{\rm{H2}} = \htwomass$ of H$_2$ in the SMC, equal to $\sim 10\%$ of
  the \hi\ mass. This is lower than the ratio of molecular to atomic gas found
  in the inner parts of spiral galaxies, but similar to that found in dwarf
  galaxies and the outer parts of spirals \citep[][and see Figure
  \ref{PRESS}]{YOUNG91,ISRAEL97B,LEROY05}.

\item The derived CO-to-H$_2$ conversion factor over the entire SMC is
  $\intxco$ \xcounits, similar to that obtained by \citet{ISRAEL97}. Toward
  molecular peaks, correcting for H$_2$ beyond the CO along the line of sight,
  we find a conversion factor of $\peakxco$ \xcounits. This represents our
  best value of \xco\ for comparison with virial mass measures. It is still
  $2$ -- $4$ times larger than virial \xco\ measurements for SMC GMCs.


\item SMC GMCs appear to be gravitationally bound. We measure M$_{\rm{H2}}$ to
  be larger than M$_{vir}$, surface densities similar to those of Milky Way
  GMCs, and a mean midplane pressure to similar to that at the Solar Circle
  (albeit with a large systematic uncertainty).

\item We find evidence for the selective photodissociation of CO. About the
  molecular peaks, H$_2$ is more extended than CO by about 30\%. Averaging
  over low CO ($I_{CO} < 0.3$ K km s$^{-1}$) and high CO ($I_{CO} > 0.3$ K km
  s$^{-1}$) regions, we find the low CO regions have twice as much H$_2$ per
  CO as the high CO regions. We average over a large region to make these
  measurements, so this is not a sensitivity effect. Regions of high
  H$_2$-to-CO occur near intense radiation fields surrounding bright CO
  clouds, consistent with the scenario in which the outer parts of SMC clouds
  see their CO destroyed by UV photons while H$_2$ survives due to
  self-shielding \citep{MALONEY88}.

\item Several clouds away from massive star formation show more CO relative to
  FIR than average, making them good candidates to harbor populations of cold
  dust. One of these clouds, SMC~B1-1 \citep{RUBIO04}, shows a greater
  submillimeter-to-FIR ratio than the SMC as a whole. These clouds represent
  $\sim 30\%$ of the total CO luminosity.

\item GMCs in the SMC are either translucent or very small. SMC molecular
  peaks have H$_2$ surface densities similar to Milky Way GMCs,
  $\Sigma_{\rm{H2}} \approx 100$ -- $200$ M$_{\odot}$ pc$^{-2}$ and
  extinctions of $A_V \approx 1$ -- $2$ mag. This is lower than lower than the
  $\bar{A_V} = 4$ -- $8$ mag found in Milky Way clouds \citep{MCKEE89}. Either
  photoionization by far-UV photons does not regulate the structure of SMC
  GMCs and they indeed have lower average extinctions or the star-forming
  parts of SMC GMCs are a factor of $\sim 3$ smaller than our 46 pc beam.

\item The low H$_2$-to-\hi\ ratio in the SMC results from a combination of low
  gas densities and metallicity effects. The SMC has a lower ratio of
  molecular to atomic gas by a factor $2$ -- $3$ than expected based on its
  hydrostatic pressure \citep[and][]{BLITZ06B}, but also a median pressure a
  factor of $\sim 3$ lower than large galaxies. The median $P_{h}$ is
  comparable to that at the Solar Circle, though very uncertain. 

\end{enumerate}
\acknowledgements This work is based on observations made with the {\em
  Spitzer Space Telescope}, which is operated by the Jet Propulsion
Laboratory, California Institute of Technology under a contract with NASA.
This research was partially supported by NSF grant AST-0228963. Partial
support for this work was also provided by NASA through an award issued by
JPL/Caltech (NASA-JPL Spitzer grant 1264151 awarded to Cycle 1 project 3316).
We made use of the NASA/IPAC Extragalactic Database (NED) which is operated by
the Jet Propulsion Laboratory, California Institute of Technology, under
contract with the National Aeronautics and Space Administration; and NASA's
Astrophysics Data System (ADS). This research has made use of the NASA/IPAC
Infrared Science Archive, which is operated by the Jet Propulsion Laboratory,
California Institute of Technology, under contract with the National
Aeronautics and Space Administration. We wish to especially thank L. Blitz, C.
McKee, and J. Graham, who all gave critical readings of this work while it was
being prepared as part of AL's thesis and F. Boulanger, who kindly provided
suggestions on reading a draft. We also thank the anonymous referee for
suggestions that improved this work.

\begin{appendix}

\section{Calibrating the MIPS 160 $\mu$m Map}
\label{CALMIPSAPP}

Figure \ref{INTSED} and Table \ref{FLUXTAB} show that there is good
photometric consistency among our data. Fluxes at the same wavelength
agree within the uncertainties, and the data is consistent with DIRBE,
which represents the best measurements of absolute sky brightness at
FIR wavelengths. The exception to the general good agreement is that
the 160 $\mu$m contains more flux than we expect based on comparison
with the DIRBE data at 140 $\mu$m and 240 $\mu$m and the ISO 170
$\mu$m map. In this section we describe how we use the DIRBE data to
calibrate the MIPS 160 $\mu$m map. The calibrated 160 $\mu$m map is
used for all of the results in this paper, including Figure
\ref{INTSED} and Table \ref{FLUXTAB}. Our approach to calibrating the
160 $\mu$m map is:

\begin{enumerate}
\item We adopt DIRBE as our absolute calibration.

\item From the DIRBE bands at 140 $\mu$m and 240 $\mu$m we estimate
the intensity at 160 $\mu$m by interpolation. We adopt $\beta = 2$ and
calculate a temperature from the ratio of the 140 $\mu$m to 240 $\mu$m
bands. The method of interpolation makes little ($\lesssim 10$)
difference because the 140 $\mu$m DIRBE band is close to the 160
$\mu$m MIPS band.

\item The foreground-subtracted DIRBE data (see \S \ref{DATA}) are our
estimate the 160$\mu$m intensity from the SMC along each line of
sight. We add our best estimate of the MIPS 160 $\mu$m foreground
(from zodiacal light, Galactic cirrus, and the CIB) to this
value. This is our best estimate of the intensity that MIPS {\em
should} measure.

\item We make the decision to model the difference between MIPS and
DIRBE using one parameter: the best fit gain. This is the ratio of the
intensities measured by MIPS to those that it should measured as
estimated from DIRBE, $I_{DIRBE,160\mu m}/I_{MIPS,160\mu m}$.

\item We derive the best value for the gain by taking the average
ratio of predicted to measured intensity over the MIPS map. We perform
the comparison at the DIRBE resolution, $0^{\circ}.7$. We use DIRBE to
fill in empty parts of the MIPS map, but we only considering lines of
sight for which at least 75\% of the data within the beam come from
MIPS. This average ratio is $0.8$, so that $I_{DIRBE,160\mu m} =
0.8~I_{MIPS,160\mu m}$; the median and integrated ratios within a few
percent of this value.

\item We apply this factor to the MIPS data before foreground
subtraction and proceed as described above. Scaling by $0.8$ before
foreground subtraction lowers the flux of the MIPS map by $\sim 5000$
Jy.
\end{enumerate}

The [CII] 158 $\mu$m line lies in the MIPS 160 $\mu$m bandpass but not
the nearby DIRBE bands. However, we show below that reasonable
estimates of the contribution by this interstellar cooling line are
too low to explain the excess flux in the MIPS map. Instead, the
difference may be a result of the mapping method or simply an offset
between the MIPS and DIRBE calibrations for the run over which the SMC
data were taken. Throughout this paper we adopt the DIRBE absolute
flux calibration, and use the rescaled MIPS map. We mention the effect
that excess emission at 160 $\mu$m would have on our results in the
discussion of uncertainties.

\begin{figure}
\begin{center}
\epsscale{0.5} 
\plotone{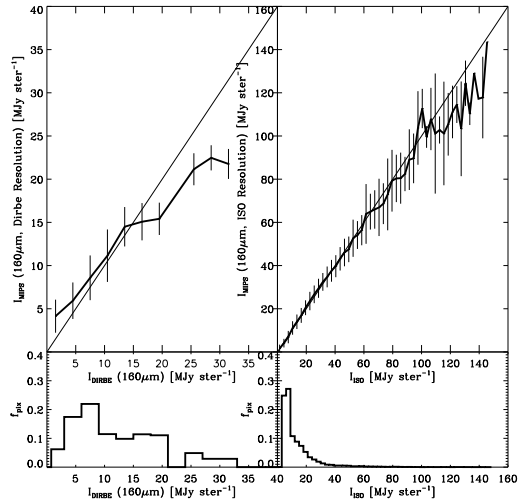}
 
\figcaption{\label{MIPSVSDIRBE} The MIPS 160 $\mu$m flux after
correction (scaling by $0.8$) and convolution to the DIRBE (left) and
ISO (right) resolution plotted as a function of (left) DIRBE
interpolated to 160 $\mu$m from nearby wavebands and (right) ISO. We
bin the data by intensity and then plot the mean and RMS scatter in
each bin. The agreement with ISO is good, but at high intensities the
MIPS data is 20 -- 30\% lower than the DIRBE data, probably because of
saturation effects. Normalized histograms show the distribution of
intensities for both data sets.}

\end{center}
\end{figure}

Figure \ref{MIPSVSDIRBE} shows the calibrated MIPS 160 $\mu$m data
convolved to $0^{\circ}.7$ and $1\arcmin.5$ and plotted against data
from DIRBE and ISO. We bin the data by intensity and plot the mean and
RMS scatter for each bin. Arbitrarily normalized histograms below each
plot show the relative distribution of intensities for ISO and DIRBE
and so give an idea of the importance for each point. The agreement
between MIPS and the other data is good, particularly over the range
of common intensities (high values in the histogram). The calibration
could be improved by adding additional parameters (e.g. an intercept
or quadratic term), but this runs the risk of skewing our results by
overprocessing the data. Figure \ref{MIPSVSDIRBE} may serve as a guide
to the reader of the uncertainty in the calibration.

In the remainder of this appendix, we show that the discrepancy between MIPS
and DIRBE is too large to be accounted for by the [CII] 158 $\mu$m fine
structure line or the known MIPS light leak.

\subsection{Is the Excess Due to [CII] Emission?} 

The [CII] 158 $\mu$m line lies in the MIPS 160 $\mu$m bandpass. Does
contamination by this important interstellar cooling line explain the
discrepancy between the MIPS and DIRBE data? No extensive maps of the [CII]
line in the SMC are available, so we estimate the [CII] contamination of the
160 $\mu$m map in two ways: by extrapolating from measurements of [CII] in a
few star-forming regions and by using the [CII]-to-FIR and [CII]-to-CO ratio
measured in other galaxies.

\citet{ISRAEL93} and \citet{BOLATTO00} both measured the [CII] line
towards several star-forming regions. Using the Kuiper Airborne
Observatory (KAO), \citet{ISRAEL93} found an integrated [CII] line
flux across five of the brightest star-forming regions that
corresponds to $\sim 50$ Jy averaged across the $4 \times 10^{11}$ Hz
MIPS bandpass (F. Israel, private communication).  Although the KAO
measurements cover only a small portion of the SMC, they do include
the brightest star-forming regions which account for $\approx 10\%$ of
the FIR emission. If [CII] emission is directly proportional to FIR
emission across the SMC (rather than preferentially concentrated in
PDR/star-forming regions) then [CII] emission from the SMC totals
$\sim 500$ Jy.

\citet{STACEY91} found that the [CII] emission accounts for $0.1$ --
$1$ \% of the FIR luminosity from a star-forming galaxy. The
integrated FIR flux from the SMC based on IRAS is $\sim 4 \times
10^{-10}$ W m$^{-2}$ so we expect $\sim10^{2}$ -- $10^{3}$ Jy from the
[CII] line in the SMC based on the FIR emission. \citet{MOCHIZUKI98}
and \citet{MADDEN97} report ratios between the [CII] and CO
1$\rightarrow$0 line fluxes of $10^4$ -- $10^5$ for some actively star
forming dwarf galaxies. \citet{MIZUNO01} finds the integrated CO
intensity of the SMC to be $\sim 10^5$ K km s$^{-1}$ pc$^2$, implying
a flux from the CO line of $3 \times 10^{-17}$ W m$^{-2}$. If the SMC
has a ratio of [CII] to CO emission consistent with the other dwarfs,
this implies a [CII] luminosity of $10^{2}$ -- $10^{3}$ Jy, consistent
with the estimate from the FIR.

Thus, several lines of argument suggest that the [CII] line
contributes $10^2$ -- $10^3$ Jy to the MIPS 160 $\mu$m band. The most
likely scenario is that a few hundred Jy of emission of total
luminosity in the MIPS bandpass can be attributed to the [CII]
line. This is significantly less than the $\sim 5000$ Jy needed to
explain the discrepancy between the MIPS 160 $\mu$m band and the
emission in nearby DIRBE bands.

\subsection{Is the Excess due to the MIPS Light Leak?} 

The known light leak into MIPS from the near-IR cannot be the source of the
flux discrepancy and does not represent a concern. According to the MIPS Data
Handbook, data with a 160 $\mu$m / 2 $\mu$m flux density ratio greater than
0.004 will be uncorrupted by the light leak. Estimated from the DIRBE data,
the median ratio of 160 $\mu$m flux density to 2 $\mu$m flux density along
lines of sight towards the SMC is $75$ and is never less than $35$.

\section{Producing the Dust Map}
\label{DUSTMAPAPP}

In this section, we describe how we use the IRAS 100 $\mu$m and MIPS
160 $\mu$m data to construct a map of the dust mass surface density,
$\Sigma_{dust}$, in the SMC. We then give brief descriptions of the
sources of uncertainty that may affect this map.

\subsection{Method}

Deriving the dust mass surface density $\Sigma_{dust}$ follows this
basic method:

\begin{enumerate}
\item Estimate the dust temperature, $T$, or distribution of
temperatures the line of sight.

\item Adopt an emissivity for the dust. This consists of a mass
absorption coefficient, $\kappa_{\nu}$, and a power law wavelength
dependence, $\kappa_{\nu} \propto \lambda^{-\beta}$, with an index
$\beta$.

\item Convert the amplitude of the FIR emission into a dust mass
surface density. For a single dust population:

\begin{equation}
\label{DUSTMASSEQN}
\Sigma_{Dust}~[\mbox{g cm}^{-2}] = \frac{I_{\nu}~[\mbox{MJy
      ster}^{-1}]}{\kappa_{\nu}~[\mbox{cm}^2 \mbox{ g}^{-1}]~B_{\nu}
      (T)~[\mbox{MJy ster}^{-1}]}~\mbox{,}
\end{equation}

\noindent where $B_{\nu}~(T)$ is the specific intensity of a blackbody
with temperature $T$ at a frequency $\nu$ and $I_{\nu}$ is the
intensity of measured FIR emission at the same frequency.
\end{enumerate}

Our data have limited spatial resolution ($\sim 75$~pc at 100$\mu$m),
and each line of sight through the SMC is probably several kiloparsecs
long with more than one \hi\ velocity component. As a result, we
expected several populations of dust to contribute along each line of
sight. Therefore, we treat the SMC using models presented by
\citet{DALE01} and refined in \citet{DALE02}. These models consider
dust that is illuminated by a distribution of radiation fields, $U$,
ranging from $0.3$ to $10^5$ times the local intensity. They use an
updated version of the description of interstellar dust by
\citet[][]{DESERT90}. The free parameters in the model are the power
law index, $\alpha$, of dust mass as a function of illuminating
radiation field and the total dust mass, $M_{dust}$, proportional to
the amplitude of the spectrum. These parameters, $\alpha$ and
$M_{dust}$, are related by

\begin{equation}
dM_{dust} \propto U^{-\alpha}~dU~.
\end{equation}

\noindent So that the lower the value of $\alpha$, the hotter the
dust, on average.

We derive $\Sigma_{dust}$ for each line of sight by using the observed
$100$-to-$160$ $\mu$m color to pick the best \citet{DALE02} model,
identified by its power law index $\alpha$. From the amplitude of the
FIR emission and the model SED associated with that $\alpha$ we
compute the dust mass along the line of sight. \citet{DALE01}
emphasize that the models are well distinguished by the
$I_{60}/I_{100}$ color alone. The 100-to-160$\mu$m color appears to do
almost as good a job while avoiding the 24 -- 70 $\mu$m bands. These
bands introduce additional free parameters because they contain
substantial contributions from polycyclic aromatic hydrocarbons (PAHs)
and very small grains (VSGs) \citep{DESERT90}, the relative mix of
which (i.e. the exact dust size distribution) is not well known in the
SMC.

\subsection{Uncertainties}

{\em Statistical}: The statistical uncertainty associated with the map is
modest. We repeatedly add noise with a magnitude from Table \ref{NOISETAB} to
the data and estimate the $1\sigma$ uncertainty in $\Sigma_{dust}$ to be 8\%
at $2\arcmin.6$ resolution and 4\% at $4\arcmin$ resolution. The uncertainty
in $M_{dust}$ is thus dominated by the choice of method, model, and systematic
uncertainties in the data.

{\em Foreground Subtraction}: The uncertainty associated with the foreground
subtraction can have a sizeable impact, particularly in the diffuse (low
intensity) Wing. We test the effect of errors in the foreground subtraction by
adding and then subtracting 1 MJy ster$^{-1}$ to the 100 $\mu$m map then the
160 $\mu$m map. This offset results in a $\pm 20\%$ change in the total dust
mass over the SMC. In regions of low intensity an offset of 1 MJy ster$^{-1}$
in one band may affect the dust surface density by up to 50\%.

{\em Calibration and Saturation}: If the MIPS calibration (rather than DIRBE)
is correct then our MIPS 160 $\mu$m map should be scaled by $\sim 1.25$. This
results in a dust content $\sim 80\%$ higher than we find. The possible
saturation concerns for regions with $I_{160} > 50$~MJy ster$^{-1}$ only have
a small impact ($< 10\%$) on the derived dust mass.

\begin{figure}
\begin{center}
\epsscale{0.75} \plottwo{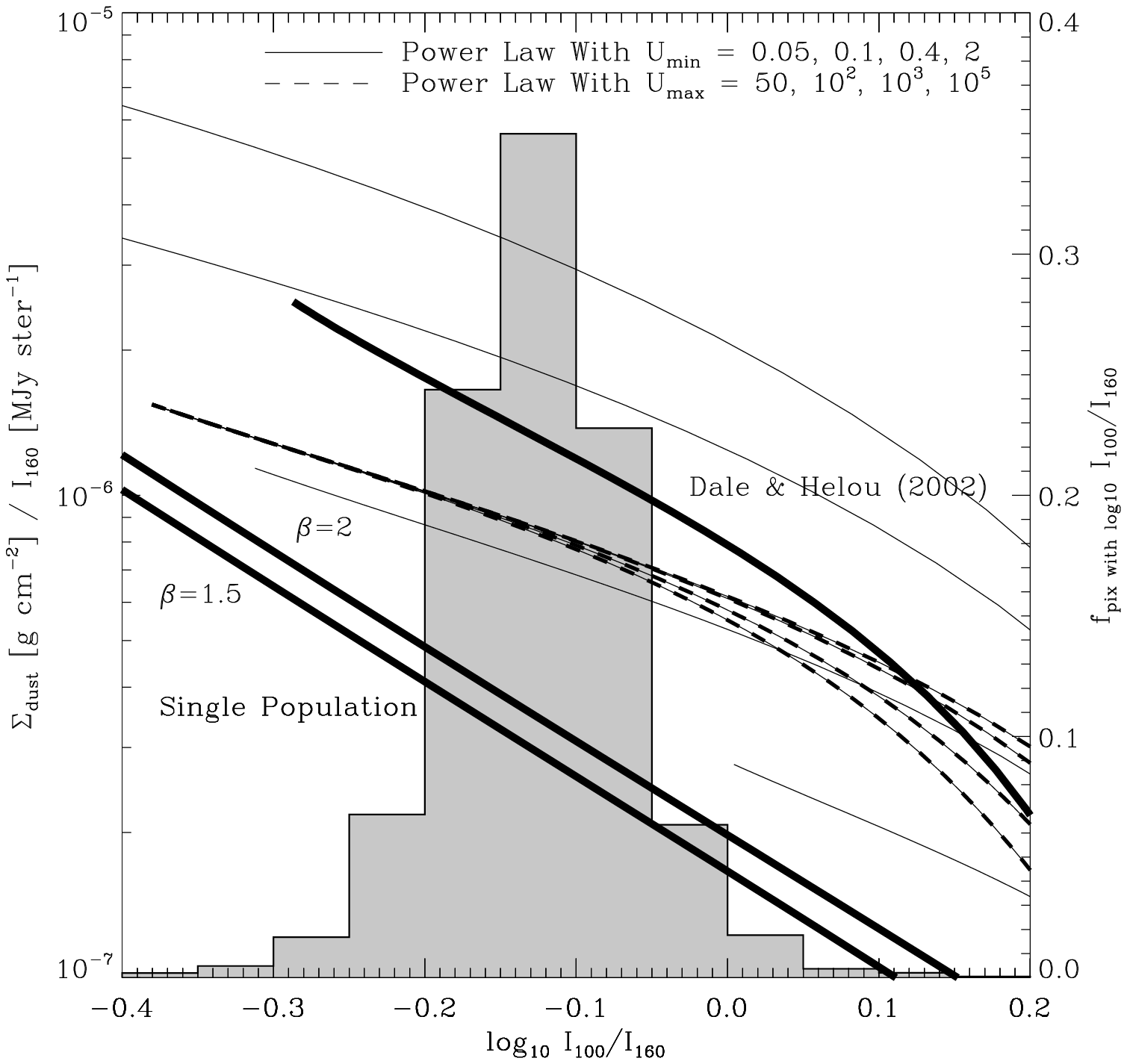}{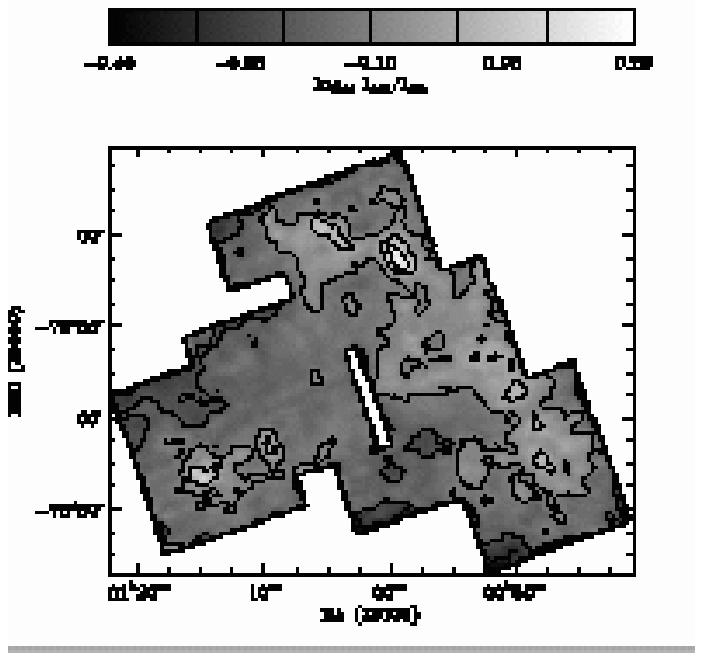}
 
\figcaption{\label{DALEMETHOD} A detailed look at the methodology used
to create our dust map. (left) A normalized histogram of the
$100$-to-$160$ $\mu$m color for the SMC along with the dust mass
surface density per unit 160 $\mu$m intensity,
$\Sigma_{dust}/I_{160}$, as a function of FIR color, $\log_{10}
I_{100}/I_{160}$, for several cases. The three bold lines show the
relationship for a single population of dust with $\beta = 1.5$, one
with $\beta = 2$, and the power law models by \citet{DALE02}. The thin
lines show the same relationship for simple power law models with the
minimum (solid) and maximum (dashed) radiation field varied. These
lines show the impact of the assumptions that go into the
\citet{DALE02} model. The maximum radiation field has very little
impact, but the minimum radiation field affects the conversion between
FIR intensity and dust mass surface density. (right) A map of the FIR
color, $\log_{10} I_{100}/I_{160}$, in the SMC. The star-forming
regions have high 100-to-160$\mu$m ratios and are easily identified
from the map.}

\end{center}
\end{figure}

{\em Power Law vs. Single Population}: The dust mass derived using the
\citet{DALE02} models is a factor of $\sim 3$ higher than one would derive
using a single $\beta = 2$ population and the 100 and 160 $\mu$m bands to
model each line of sight. Figure \ref{DALEMETHOD} shows this difference by
plotting the key quantity, $\Sigma_{dust}$ per unit 160 $\mu$m emission, as a
function of the 100-to-160 $\mu$m color. Figure \ref{DALEMETHOD} also shows
the relative distribution of sight-lines with each color in the SMC and
illustrates several other systematic effects discussed in this section.

{\em $U_{min}$ and $U_{max}$}: The upper and lower limits to the radiation
field distribution, $U_{min}$ and $U_{max}$, may affect the derived dust mass.
\citet{DALE01,DALE02} constructed their models to match the integrated SEDs of
a sample of several dozen star-forming galaxies, so $U_{min} = 0.3$ and
$U_{max} = 10^5$ times the local value are motivated by observations. However,
we test the effects of varying $U_{min}$ and $U_{max}$ on a simple model that
includes only big grains with $\kappa_{250} = 8.5$~cm$^{2}$~g$^{-1}$. If
$U_{min} = 0.1$ is adopted instead of $0.3$ then $M_{dust}$ will be $\sim 2$
times higher. If $U_{min} = 0.05$ instead of $0.3$ then $M_{dust}$ will be a
factor of $4$ higher. Higher values of $U_{min}$ imply less dust mass, but
$U_{min}$ can not be much higher than $\approx 0.5$ or the models will not be
able to reproduce the lowest $I_{100}/I_{160}$ ratios seen in the SMC. Varying
$U_{max}$ has minimal effect over the range of colors observed in the SMC.
Dust illuminated by $U = 0.3$ has $T \approx 14.5$~K, so the uncertainty
regarding $U_{min}$ is a restatement of the potential population of very cold
dust. Dust with $T \lesssim 15$~K has $log_{10} I_{100}/I_{160} \lesssim
-0.5$. Figure \ref{DALEMETHOD} shows that there are very few lines of sight in
the SMC with $log_{10} I_{100}/I_{160} \lesssim -0.3$, let along $-0.5$.

{\em Choice of Bands (Inclusion of VSGs)}: The \citet{DALE02} models account
for the contribution from VSGs, albeit with a Galactic mix of grain types.
Therefore, the choice to omit the 60 $\mu$m and 70 $\mu$m bands has a
relatively small impact. Using the 100 and 160 $\mu$m data yields a mass 50 \%
higher than the 60 and 100 $\mu$m data, presumably because the SMC has more
VSGs than the Milky Way mix adopted for the model. For the case of a single
dust population it is critical to account for emission from VSGs if shorter
wavelengths are included; failing to do so may result in estimates of dust
masses an order of magnitude too low \citep[this explains a factor of $\sim 5$
out of the difference between our results and those of][the choice of a power
law as opposed to a single model explains the
balance]{SCHWERING88,STANIMIROVIC00}.

{\em Mass Absorption Coefficient ($\kappa_{\nu}$)}: \citet{ALTON04} synthesize
determinations of the emissivity from comparisons of optical and sub-mm/FIR
emission along the same line of sight. They show that determinations of
$\kappa_{250}$ span from $\approx 5$ cm$^2$ g$^{-1}$ to $15$ cm$^2$ g$^{-1}$,
a substantial scatter. The \citet{DALE02} models use a modified version of the
\citet{DESERT90} dust model, which in turn adopts a slightly modified version
of the \citet{DRAINE84} long wavelength mass absorption coefficient. Therefore
the dust map in this paper adopts $\kappa_{250} \approx 8.5$ cm$^2$ g$^{-1}$
\citep{DRAINE84}, a widely used value close to the middle of the range of
literature determinations. The dust surface densities we derive are uncertain
by a factor of $2$ due to the uncertainty in the mass absorption coefficient.
All of these values are not equally likely, but the value of $\kappa_{250}
\approx 5$ cm$^2$ g$^{-1}$ \citep{LI01} is also commonly used.

{\em Emissivity Dependence on Wavelength ($\beta$)}: At long wavelengths, the
wavelength dependence of $\kappa_{\nu}$ is usually treated as a power law with
an index $\beta$, so that $\kappa_{\nu} \propto \lambda^{-\beta}$. However,
there is substantial evidence that $\beta$ may vary with environment
\citep{DUPAC03,ALTON04}. The models by \citet{DALE02} incorporate an empirical
radiation field dependence for $\beta$ calibrated using sub-mm data for the
sample of galaxies used to develop the \citet{DALE01,DALE02} models. For this
calibration $\beta = 2.5$, $2.1,$ and $1.3$ for radiations fields $1,$ $10,$
and $1000$ times the local value. For the values of $\alpha$ we find in the
SMC, $\beta \approx 2$. This may be too high, $\beta = 1.5$ may be more
appropriate (see \S \ref{INTSEDSECT}). Because the 100 and 160 $\mu$m bands
are fairly close in wavelength, the impact of changing $\beta$ by this amount
is small; we would find $M_{dust}$ to be $\approx 20\%$ lower for $\beta =
1.5$ than $\beta = 2$. Though it has a small effect on $M_{dust}$, $\beta$ is
important to interpreting the sub-mm excess observed towards the SMC
\citep{AGUIRRE03}. Further, since $\beta$ seems to be a function of
environment, variations between diffuse and dense gas may systematically
affect our H$_2$ map (see main text).

\begin{deluxetable}{l c l}
\tabletypesize{\small}
\tablewidth{0pt}
\tablecolumns{5}
\tablecaption{\label{DUSTUCTAB} Uncertainties in the Dust Map}

\tablehead{\colhead{Description} & \colhead{Sense} &
  \colhead{Magnitude} }

\startdata
Data & & \\
\ldots Statistical\tablenotemark{a} & $\pm$ & $\lesssim 10\%$ \\
\ldots Foreground & $\pm$ & $20\%$ \\
\ldots Calibration & $+$ & $80\%$ \\
\ldots Saturation & $+$ & $\lesssim 10\%$ \\
\hline
Choice of Model & & \\
\ldots Power Law vs. Single Pop & divide & $3$ \\
\ldots Choice of bands\tablenotemark{b} & - & $50\%$ \\
\ldots $U_{min}$\tablenotemark{c} & multiply & $2+$ \\
\ldots $\kappa_{250}$ & multiply/divide & $2$ \\
\ldots $\beta$ & $-$ & $20\%$
\enddata
\tablenotetext{a}{Per line of sight. Negligible over the whole SMC.}
\tablenotetext{b}{If the 60 and 100 $\mu$m bands are used instead of
  the 100 and 160 $\mu$m bands.}
\tablenotetext{c}{In the limit, this is the same as the cold dust concern.}
\end{deluxetable}

\end{appendix}

\end{document}